\pdfoutput=1

\documentclass{aastex6}
\newcommand{\DDt}[1]{\frac{\mathrm{D} #1}{\mathrm{D} t}}

\usepackage{amsmath}

\usepackage[all]{hypcap}

\begin{document}

\title{Temperature Structure and Atmospheric Circulation of 
 Dry, Tidally Locked Rocky Exoplanets}
\author{Daniel D.~B.~Koll and Dorian S.~Abbot}
\affil{Department of the Geophysical Sciences, University of Chicago,
    Chicago, IL 60637}
\email{dkoll@uchicago.edu}

\begin{abstract}
  Next-generation space telescopes will observe the atmospheres of
  rocky planets orbiting nearby M-dwarfs. Understanding these
  observations will require well-developed theory in addition to
  numerical simulations. Here we present theoretical models for the
  temperature structure and atmospheric circulation of dry, tidally
  locked rocky exoplanets with grey radiative transfer and test them
  using a general circulation model (GCM).  First, we develop a
  radiative-convective model that captures surface temperatures of
  slowly rotating and cool atmospheres. Second, we show that the
  atmospheric circulation acts as a global heat engine, which places
  strong constraints on large-scale wind speeds. Third, we develop a
  radiative-convective-subsiding model which extends our
  radiative-convective model to hot and thin atmospheres.  We find
  that rocky planets develop large day-night temperature gradients at
  a ratio of wave-to-radiative timescales up to two orders of
  magnitude smaller than the value suggested by work on hot Jupiters.
  The small ratio is due to the heat engine inefficiency and asymmetry
  between updrafts and subsidence in convecting atmospheres.  Fourth,
  we show using GCM simulations that rotation only has a strong effect
  on temperature structure if the atmosphere is hot or thin.  Our
  models let us map out atmospheric scenarios for planets such as GJ
  1132b and show how thermal phase curves could constrain
  them. Measuring phase curves of short-period planets will require
  similar amounts of time on the \textit{James Webb Space Telescope}
  as detecting molecules via transit spectroscopy, so future
  observations should pursue both techniques.
\end{abstract}

\section{Introduction}
\label{sec:introduction}

\subsection{Importance of atmospheric dynamics}

Terrestrial exoplanets orbiting M-dwarfs are extremely common. Results
from the \textit{Kepler} space telescope show that there are at least
$\sim0.5$ rocky planets per M-dwarf, half of which could even be
habitable \citep{dressing2015}. Just as important, near-future
telescopes like the \textit{James Webb Space Telescope}
(\textit{JWST}) will be able to characterize the atmospheres of these
planets \citep{deming2009c,beichman2014,cowan2015}, making them one of
the most promising observational targets of the coming decade.

New theories are needed to understand the potential atmospheres of
these exoplanets, particularly their temperature structures and
large-scale circulations.  An atmosphere's temperature structure and
circulation critically influence a planet's surface and atmospheric
evolution as well as its potential habitability \citep{kasting1988,abe2011,yang2013}.
An atmosphere's temperature structure and circulation are also
important for interpreting observations.
For example, a planet's
emission spectrum is determined by the vertical temperature distribution
of its atmosphere, while the planet's thermal and optical phase curves
are set by its day-night temperature gradient and cloud patterns
\citep{seager2009,yang2013,hu2015}. Even transit measurements can be
strongly influenced by chemical mixing and clouds, which in turn
depend on the atmosphere's large-scale circulation
\citep{fortney2005,parmentier2013,charnay2015b,line2016}.

Unfortunately there is a large gap between current theories of
terrestrial atmospheres and the wide range of potential
exoplanets. Planets accessible to follow-up observations will
generally be in short-period orbits, experience strong tidal forces,
and thus tend to be either tidally locked or captured in higher
spin-orbit resonances \citep{kasting1993a,makarov2012a}.  The solar
system offers no direct analogs of such atmospheres, and their
dynamics are still poorly understood.
In this work we focus on
tidally locked (synchronously rotating) atmospheres because
  their dynamics would differ most drastically from rapidly rotating
  atmospheres, whereas planets in higher
  spin-orbit resonances should resemble hybrids between tidally locked
  and rapidly rotating planets (also see Section
  \ref{sec:discussion}).

\subsection{Previous work and open questions}
\label{sec:previous}

Many groups have already used general circulation models (GCMs) to study
the thermal structure and atmospheric circulation of tidally locked
terrestrial planets
\citep[e.g.,][]{joshi1997,merlis2010,heng2011a,pierrehumbert2011c,selsis2011,leconte2013,yang2013,zalucha2013,wordsworth2015,kopparapu2016}.
These studies investigated a range of processes that shape the
atmospheres of tidally locked planets, including the large-scale
day-night circulation, equatorial superrotation, heat transport by
atmospheric waves, and the potential for atmospheric collapse if the
nightside becomes too cold.
The development of theory to understand these
processes, however, has not kept up with the rapid proliferation of simulations.

Recent theories of rocky planets focused on planets for
which the horizontal heat redistribution is extremely efficient
\citep{pierrehumbert2011c,mills2013,yang2014,wordsworth2015}.
Among the latter, \citet{pierrehumbert2011b} developed a scaling
relation for the surface temperature of a planet that is
horizontally completely uniform and whose atmosphere is optically thick,
\begin{eqnarray}
  T_{s} & = T_{eq} \times
                                        \dfrac{\tau_{LW}^{\beta}}{\Gamma\left(1+4\beta\right)^{1/4}}.
\label{eqn:pierrehumbert1}
\end{eqnarray}
Here $T_{eq}$ is the planet's equilibrium temperature defined as
$T_{eq} \equiv [L_* (1-\alpha)/(4\sigma)]^{1/4}$, $\tau_{LW}$ is
the longwave optical thickness, $\Gamma$ is the Gamma function
defined as $\Gamma(a) \equiv \int_0^{\infty} t^{a-1} \exp(-t) dt$, and
$\beta\equiv R/(c_p n)$. $L_*$ is the stellar constant, $\alpha$ is the
planetary bond albedo, $\sigma$ is the Stefan-Boltzmann constant, $R$
is the specific gas constant, $c_p$ is the specific heat capacity, and
$n$ governs how optical thickness depends on pressure
(Section \ref{sec:methods}).
Similarly, \citet{wordsworth2015} developed a theory for the
temperature structure of tidally locked atmospheres in the
optically thin limit, in which atmospheres become particularly
vulnerable to atmospheric collapse.
\citet{wordsworth2015} found a lower
bound for the nightside temperature of a tidally locked
planet\footnote{We use Equation \ref{eqn:wordsworth1} instead of Equation (29) in
  \citet{wordsworth2015}, because it does not assume a specific value for $n$.},
\begin{eqnarray}
  T_{n} & = T_{eq} \left(\dfrac{\tau_{LW}}{2}\right)^{1/4}.
\label{eqn:wordsworth1}
\end{eqnarray}
Common to both scalings is that they are not valid in
the physically important regime of optical depth unity, and indeed
contradict each other when extrapolated to this limit.
Neither do they explicitly account for
horizontal atmospheric dynamics\footnote{\citet{wordsworth2015} also
  developed a model that incorporates dynamics, which we revisit
  in Section \ref{sec:winds}.}. Nevertheless, we expect that the
dynamics of tidally locked planets should be sensitive to a range
of additional processes, including the atmosphere's radiative
timescale, surface drag, and planetary rotation, all of which
have not yet been addressed for rocky exoplanets.

On a different front, recent work has begun to understand the
atmospheric circulation of hot Jupiters
\citep{perez-becker2013a,showman2015,komacek2016}.
\citet{perez-becker2013a} developed a weak-temperature-gradient (WTG)
theory that explains why the hottest hot Jupiters also tend to have
the highest day-night brightness temperature contrasts.  WTG describes
atmospheres that are slowly rotating and are relatively cool, which
allows atmospheric waves to efficiently eliminate horizontal
temperature gradients \citep{showman2013b}. In equilibrium the wave
adjustment leads to subsidence, that is, sinking motions, in regions of
radiative cooling (see Section \ref{sec:radsub}).
\citet{perez-becker2013a} showed that day-night temperature
gradients become large once
the radiative timescale becomes shorter than the timescale
for subsidence, $t_{rad} \lesssim t_{sub}$. On hot Jupiters with
sufficiently strong drag, temperature gradients are large when
\begin{eqnarray}
  t_{rad} & \lesssim & \frac{t_{wave}}{t_{drag}} \times
                       t_{wave},
\label{eqn:perez-becker1}
\end{eqnarray}
where $t_{wave}$ is the timescale for a gravity wave to
horizontally propagate across the planet and $t_{drag}$ is a
characteristic drag timescale.

It would be tempting to assume Equation \ref{eqn:perez-becker1}
applies equally well to rocky exoplanets. That is not the case, and
published GCM results of rocky planets are already at odds with it.
We show in Appendix \ref{sec:appendix1} that for most tidally locked
terrestrial planets drag and wave timescales are comparable,
$t_{drag} \approx t_{wave}$. If Equation \ref{eqn:perez-becker1} applied to
rocky planets, they should develop large day-night
temperature gradients when
\begin{eqnarray}
  1 & \lesssim & \frac{t_{wave}}{t_{rad}}.
\label{eqn:perez-becker2}
\end{eqnarray}
In contrast, the GCM simulations in \citet{selsis2011} indicate that
tidally locked rocky planets can develop atmospheric temperature
gradients at a surface pressure of about 1 bar (their Fig.~5), which
translates to a much lower value of
$t_{wave}/t_{rad}\sim0.05$. Similarly, we found in \citet{koll2015}
that rocky exoplanets develop large day-night brightness temperature
contrasts when $t_{wave}/t_{rad} \gtrsim 10^{-2}$. The
disagreement between hot Jupiter theory and rocky planets has not been
explored yet.
Here we will show that the qualitative threshold for a WTG atmosphere
to develop large temperature gradients, $t_{rad} \lesssim t_{sub}$,
also applies to rocky planets. However, rocky planets end up behaving
quite differently than hot Jupiters because of the processes that
determine the large-scale circulation and the subsidence timescale $t_{sub}$.

\subsection{Outline}
\label{sec:now}

In this paper we develop a series of models to understand the
atmospheres of tidally locked rocky exoplanets.  To show how our
models complement previous theories we adopt our nondimensional
analysis from \citet{koll2015}. Using the Buckingham-Pi theorem
\citep{buckingham1914}, we showed that the dynamics of a dry and
tidally locked atmosphere with grey radiation are governed by only six
nondimensional parameters. This set of nondimensional
parameters allows us to cleanly disentangle the
atmospheric processes that need to be addressed. One choice
for the six parameters is given by
\begin{equation}
  \left( \frac{R}{c_p}, \frac{a^2}{L_{Ro}^2},
    \frac{t_{wave}}{t_{rad}},\tau_{SW},\tau_{LW},
    \frac{t_{wave}}{t_{drag}} \right).
  \label{eqn:params}
\end{equation}
The convective lapse rate is controlled by $R/c_p$. The nondimensional
Rossby radius $a^2/L_{Ro}^2$ governs the influence of planetary
rotation on equatorial waves. Here $a$ is the planetary radius, the
equatorial Rossby deformation radius is defined as
$L_{Ro}\equiv \sqrt{a c_{wave}/(2\Omega)}$, $\Omega$ is the planetary
rotation rate, and $c_{wave}$ is the speed of a gravity wave. Although
$c_{wave}$ is a priori unknown, because it depends on an atmosphere's
vertical temperature structure, we can place a reasonable upper bound
on it by assuming an isothermal atmosphere. This assumption leads to
$c_{wave}=\sqrt{R/c_p} \times \sqrt{g H} = \sqrt{R/c_p} \times \sqrt{R
  T_{eq}}$,
where $g$ is the acceleration of gravity and $H\equiv R T_{eq}/g$ is
the scale height. The wave-to-radiative timescale ratio,
$t_{wave}/t_{rad}$, compares the time it takes for equatorial waves to
redistribute energy across the planet, $t_{wave} \equiv a/c_{wave}$,
to the atmosphere's radiative cooling time,
$t_{rad} \equiv c_p p_s/(g \sigma T_{eq}^3)$. The atmospheric
shortwave and longwave optical thicknesses are $\tau_{SW}$ and
$\tau_{LW}$. The ratio of wave to drag timescales
$t_{wave}/t_{drag} = C_D a/H$ governs surface friction and turbulent
heat fluxes (Appendix \ref{sec:appendix1}).

We only consider atmospheres that are transparent to shortwave
absorption ($\tau_{SW}=0$), which ensures that the solid surface
  substantially affects the atmospheric dynamics. This means we
  exclude from our consideration potential ``rocky'' planet scenarios
with a bulk silicate composition, but with gaseous envelopes several
hundreds of bar thick \citep{owen2016}. We expect that the observable
atmospheres of such planets would resemble gas giants more than rocky
planets, with dynamics that are better captured by theories developed
for hot Jupiters \citep{perez-becker2013a,showman2015,komacek2016}.

Of the six nondimensional parameters $\tau_{SW}$ and $\tau_{LW}$
govern radiative transfer, $R/c_p$ sets the vertical temperature
structure, and the remaining three parameters determine the horizontal
dynamics.
As an important starting point, we formulate an analytical
radiative-convective (RC) model for the
temperature structure of tidally locked atmospheres that only depends
on $\beta \equiv R/(n c_p)$ and $\tau_{LW}$ and therefore addresses
the first two processes (Section \ref{sec:radconv}). 
In the optically thick regime this model reduces to the asymptotic
limit found by \citet{pierrehumbert2011b}.
We then turn to horizontal dynamics. We show the day-night circulation
acts as a heat engine, in which heating and cooling balance frictional
dissipation in the dayside boundary layer (Section
\ref{sec:winds}). We use our heat engine theory to develop a
radiative-convective-subsiding (RCS) model that includes the effects
of $t_{wave}/t_{drag}$ and $t_{wave}/t_{rad}$ on temperature structure
(Section \ref{sec:radsub}).  For cool/thick atmospheres the RCS model
reduces to the RC model, whereas for optically thin and hot/thin
atmospheres it reduces to the asymptotic limit found by
\citet{wordsworth2015}.  Our RCS model explains why rocky planets
develop large day-night temperature gradients at a significantly lower
$t_{wave}/t_{rad}$ threshold than hot Jupiters (Section
\ref{sec:transition}).
Next, we use GCM simulations to address rapidly rotating planets and
$a^2/L_{Ro}^2$ (Section \ref{sec:rotation}). We find that
$t_{wave}/t_{rad}$ has to be big for rotation to have a strong effect
on temperature structure, that is, cause large eastward hot spot
offsets or cold nightside vortices.
Our results imply that planets like GJ 1132b or HD 219134b will likely
have significant day-night temperature contrasts, unless their
atmospheres are dominated by H$_2$ (Section \ref{sec:obs}).  We
estimate that detecting these potential contrasts via thermal
phase curves will require about as much time with
\textit{JWST} as detecting molecular signatures via transit spectroscopy.
Finally, we discuss and summarize our results in Sections
\ref{sec:discussion} and \ref{sec:conclusions}.
Appendix \ref{sec:appendix1} contains a derivation of the
characteristic drag timescale for tidally locked rocky planets,
Appendix \ref{sec:appendix2} explains how we compute the wind speed
scaling proposed by \citet{wordsworth2015}, Appendix
  \ref{sec:appendix3} describes how we solve the RCS model, and Appendix \ref{sec:appendix4} lists the
atmospheric equations of motion and radiative transfer for reference.

\section{Methods}
\label{sec:methods}

We compare our models with a large number of GCM simulations.  We use
the FMS GCM with two-band grey gas radiative transfer and dry
(non-condensing) thermodynamics. FMS has been used to simulate the
atmospheres of Earth \citep{frierson2006}, Jupiter \citep{liu2011},
hot Jupiters \citep{heng2011a}, tidally locked terrestrial planets
\citep{merlis2010,mills2013,koll2015}, and non-synchronously rotating
terrestrial planets \citep{kaspi2015}.  We use the same FMS
configuration as \citet{koll2015}. The model version we use simulates
the full atmospheric dynamics and semi-grey (shortwave and longwave)
radiation, and we include instantaneous dry convective adjustment. Drag is
parametrized using a standard Monin-Obukhov scheme which
self-consistently computes the depth of the planetary boundary layer
as well as turbulent diffusion of heat and momentum.  The surface is
represented by an idealized ``slab layer'', that is a single layer
with uniform temperature and fixed depth. The ``slab'' temperature can
be interpreted as a temperature average across the surface's thermal
skin depth \citep{pierrehumbert2011b}.  Our simulations are all
tidally locked and orbits are assumed to be circular so that the
stellar flux is constant in time.

%%%%%
\begin{table}[t!]
\tablenum{1}
\begin{center}
  \begin{tabular}{l c c c c}
      Parameter & Symbol & Unit & Minimum value &  Maximum value \\
\tableline
      Planetary radius & $a$ & $a_{\Earth}$ & $0.5$ & $2$ \\
      Rotation rate & $\Omega$ & days$^{-1}$ & $2\pi/100$ & $2\pi/1$ \\
      Equilibrium temperature & $T_{eq}$ & K &  $250$ & $600$ \\
      Surface gravity & $g$ & 10 m s$^{-2}$ & $\frac{2}{5} \times (a/a_\Earth)$ & $\frac{5}{2} \times (a/a_\Earth)$ \\
      Specific heat capacity\tablenotemark{a} & $c_p$ & J kg$^{-1}$ K$^{-1}$ & $820$ & $14518$  \\
      Specific gas constant\tablenotemark{a} & $R$ & J kg$^{-1}$ K$^{-1}$ & $189$ & $4158$ \\
      Surface pressure & $p_s$ & bar & $10^{-2}$ & $10$ \\
      Longwave optical thickness at 1 bar & $\tau_{LW,\mathrm{1 bar}}$ & - & $0.1$ & $100$\\
      Surface drag coefficient & $C_D$, via $k_{vk}$ & - & $\times 0.1$ & $\times 10$ \\
\tableline
    \end{tabular}\\
\end{center}
    \tablenotetext{a}{Minimum values correspond to CO$_2$,
      maximum values correspond to H$_2$.}
    \caption{Parameter bounds for our simulations. The shortwave
      optical thickness is set to zero,  $\tau_{SW}=0$.
      $C_D$ is not a fixed parameter, so we vary the 
      von Karman constant $k_{vk}$ to increase and decrease $C_D$ by an order of
      magnitude (Appendix \ref{sec:appendix1}). We vary $R$ and $c_p$, but require that $R/c_p$
      stays within the range of diatomic and triatomic gases ($0.23 \leq R/c_p
      \leq 0.29$).
  \label{tab:minmax}}
\end{table}
%%%%

Because we only consider atmospheres that are transparent to shortwave
radiation ($\tau_{SW}=0$), the incoming stellar flux and the planetary
albedo are degenerate in their effect on planetary temperature. For
simplicity we set the surface albedo to zero in all our simulations
and vary the incoming stellar flux.
To specify the relation between
longwave optical thickness and pressure, we use a standard power law
of the form
\begin{eqnarray}
\frac{\tau}{\tau_{LW}} = \left(\frac{p}{p_s}\right)^n. \label{eqn:powerlaw}
\end{eqnarray}
The exponent $n$ specifies how the optical thickness $\tau$ increases
with pressure. For example, $n=1$ if the opacity of a gas mixture is
independent of pressure, and $n=2$ if the opacity increases due to
pressure broadening \citep{pierrehumbert2011b,robinson2012}. Our GCM
results assume $n=2$ or $n=1$. The longwave optical thickness
  $\tau_{LW}$ is set independently of the
  atmosphere's bulk composition. To constrain $\tau_{LW}$ we note that
  more complex radiative transfer calculations tend to find values of
  $\tau_{LW}$ beween $\sim$1 and $\sim$10 at $\sim$1 bar across a wide
  range of atmospheres, \citep{robinson2014b,wordsworth2015}. We
extend these bounds by one order of magnitude in each direction and
require that the optical thickness at 1 bar satisfy
$0.1\leq\tau_{LW,\mathrm{1 bar}}\leq100$. The parameter bounds for our
simulations are summarized in Table \ref{tab:minmax}.

%%%%
\begin{figure}[tbh!]
%\figurenum{1}
%\plotone{f1.eps}
\plotone{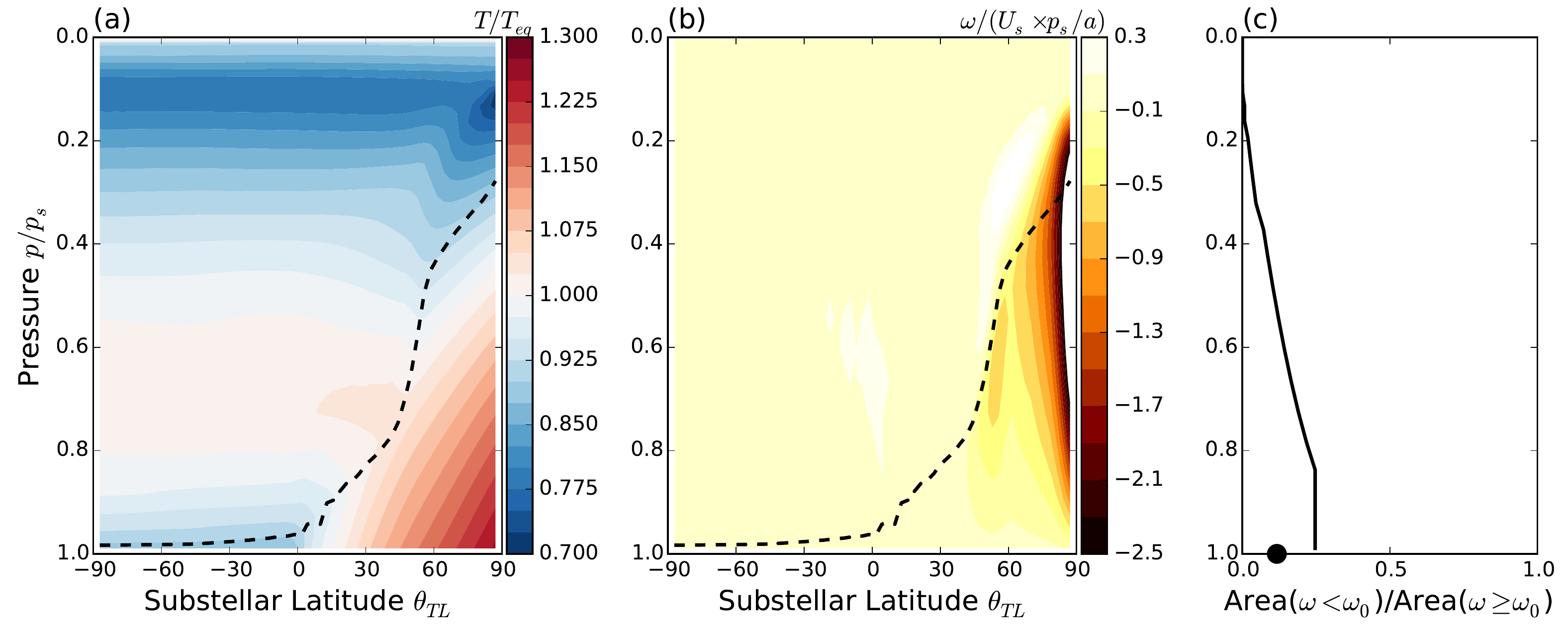}
  \caption{Temperature and circulation structure of a representative
    slowly rotating and weakly forced GCM simulation with
    $(a^2/L_{Ro}^2,t_{wave}/t_{rad})=(0.12,5\times10^{-3})$.  (a)
    Temperature as a function of substellar latitude (=0$^\circ$ at
    the terminator, =$90^{\circ}$ at the substellar point).  (b)
    Vertical velocity in pressure coordinates as a function of
    substellar latitude.  (c) Area fraction of rising motion, where
    the dot shows the vertically averaged area fraction.  The dashed
    black line in (a,b) shows the top of the GCM's boundary
    layer. Inside the boundary layer temperature increases towards the
    substellar point, and air rises; outside the boundary layer
    temperature contours are flat and air sinks. The region of
      rapidly rising motions,
      $\omega<0.01\times min(\omega)$, is narrowly focused on
      the substellar point while most of the atmosphere experiences
      weak subsidence, $\omega\gtrsim0$. We normalize temperature by
    the equilibrium temperature $T_{eq}$, and pressure velocity by the
    characteristic surface speed from the heat engine
    $U_s \times p_s/a$ (Section \ref{sec:winds}). The planet's
    physical parameters are
    $T_{eq}=283 \mathrm{K},a=a_\Earth,\Omega=2\pi/(50
    \mathrm{d}),p_s=1 \mathrm{bar},\tau_{LW}=1,$
    and $(R,c_p)=(R,c_p)_{N_2}$.
  \label{fig:climDry50a}}
\end{figure}
%%%%

Figure \ref{fig:climDry50a}a shows the temperature structure of a
representative, slowly rotating and relatively cool, GCM
simulation. The planet is Earth-sized ($a=a_\Earth$), temperate
($T_{eq}=283$K), has an orbital period and rotation rate of 50 days,
has a moderately thick $N_2$-dominated atmosphere ($p_s=1$ bar), and a
longwave optical depth of unity ($\tau_{LW}=1$). The GCM does not
explicitly model a host star, but the orbital period and equilibrium
temperature correspond to an early M-dwarf \citep[M0 or
M1;][Table 1]{kaltenegger2009}.  In terms of nondimensional
parameters,
$(R/c_p,a^2/L_{Ro}^2,t_{wave}/t_{rad},\tau_{LW},t_{wave}/t_{drag}) =
(2/7,0.12,5\times10^{-3},1,1.4)$.
Because the temperature structure is approximately symmetric about the
substellar point we present this simulation in terms of a substellar
latitude, i.e., the angle between substellar and antistellar point
\citep[also see][Appendix B]{koll2015}.  The temperature structure in
Figure \ref{fig:climDry50a}a is comparable to that found by previous
studies, and temperature contours are horizontally flat outside the
dayside boundary layer (dashed black line). The flat temperature
contours are characteristic of the weak-temperature-gradient
(WTG) regime.
However, WTG does not hold on large parts of the dayside
where the absorbed stellar flux creates a region of strong convection
and turbulent drag, which damps atmospheric waves
and allows the atmosphere to sustain horizontal
temperature gradients \citep{showman2013b}. As noted
by \citet{wordsworth2015}, this boundary layer will be of critical
importance for understanding the atmospheric circulation of rocky
planets.

\section{A two-column radiative-convective model}
\label{sec:radconv}

%%%%
\begin{figure}
\begin{center}
%\figurenum{2}
  %\plotone{f2.pdf}
  \includegraphics[width=0.75\textwidth,natwidth=975,natheight=640]{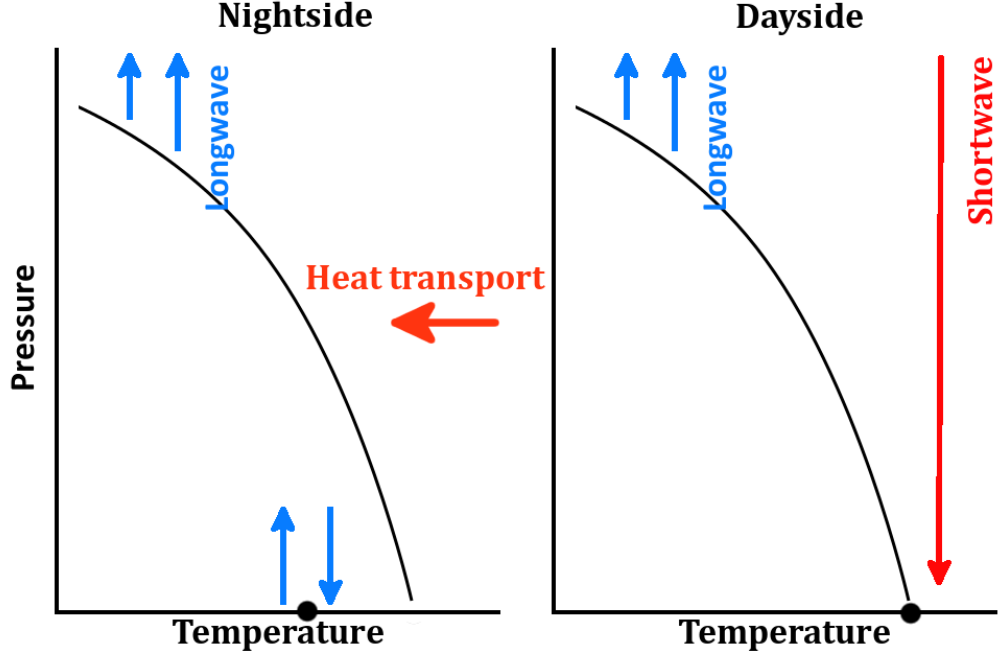}
  \caption{The two-column radiative-convective model. We assume
    convection on the dayside sets up an adiabatic temperature
    profile. Horizontal heat transport is assumed to be effective so that
    the atmosphere is horizontally uniform. The black dots indicate
    surface temperatures. The dayside surface and atmosphere are closely
    coupled via convection, whereas the nightside surface is in radiative
    equilibrium with, and generally colder than, the overlying atmosphere.
  \label{fig:diagRCmodel}}
\end{center}
\end{figure}
%%%%

In this section we present a two-column model for tidally locked
planets. We divide the planet into two (dayside and nightside)
vertical columns, as shown in Figure \ref{fig:diagRCmodel}.  The
dayside is heated by stellar radiation, which triggers convection and
sets an adiabatic vertical temperature profile.  We assume
the convective heat flux is large so that
the temperature jump between dayside surface and lowest atmospheric
level is small. We also assume convection is deep and do not
include a stratosphere (i.e., a purely radiative layer in the upper
atmosphere), so that the dayside column temperature profile in terms
of optical thickness $\tau$ can be written as
\begin{eqnarray}
  T=T_d ~ \left(\frac{\tau}{\tau_{LW}}\right)^{\beta}, \label{eqn:adiabat}
\end{eqnarray}
where $T_d$ is the dayside surface temperature, $\tau_{LW}$ is the
total optical thickness in the longwave, and $\beta \equiv R/(c_p
n)$ is the adiabatic lapse rate in optical thickness coordinates.
Next, we assume the weak-temperature-gradient (WTG) regime holds globally (i.e., also inside the dayside
boundary layer). The atmosphere is therefore
horizontally homogeneous and the nightside temperature structure is
also described by Equation \ref{eqn:adiabat}.
Under these assumptions the entire atmosphere is in
radiative-convective equilibrium, with convection governed by the
dayside surface temperature $T_d$.  The nightside surface will
generally be colder than the overlying air, which leads to stable
stratification and suppresses turbulent fluxes between the
nightside surface and atmosphere. We idealize this situation by
assuming that the nightside surface is in radiative equilibrium with
the overlying atmosphere (see Fig.~\ref{fig:diagRCmodel}).

For a grey atmosphere on a dry adiabat, the
top-of-atmosphere (TOA) upward longwave and surface downward longwave
fluxes are \citep{pierrehumbert2011b,robinson2012}
\begin{subequations}
\begin{eqnarray}
  F^{\uparrow} (\tau=0) & = & \sigma T_d^4 e^{-\tau_{LW}} + \sigma T_{d}^4 \int_0^{\tau_{LW}}
                                  \left(\frac{\tau'}{\tau_{LW}}\right)^{4\beta}
                                  e^{-\tau'} d\tau', \\
  F^{\downarrow} (\tau=\tau_{LW}) & = & \sigma T_{d}^4 \int_0^{\tau_{LW}}
                                  \left(\frac{\tau'}{\tau_{LW}}\right)^{4\beta}
                                  e^{-(\tau'-\tau_{LW})} d\tau'.
\end{eqnarray}
\end{subequations}
 Using these expressions we write the dayside TOA,
nightside TOA, and nightside surface energy budgets\footnote{We
  implicitly use the dayside surface energy budget by assuming
  that the surface-air temperature jump is negligible on the dayside.}
as
\begin{subequations}
\begin{eqnarray}
  \frac{L_*(1-\alpha_p)}{2} & = & \sigma T_{d}^4 e^{-\tau_{LW}} +
                                  \sigma T_{d}^4 \int_0^{\tau_{LW}}
                                  \left(\frac{\tau'}{\tau_{LW}}\right)^{4\beta}
                                  e^{-\tau'} d\tau' +
                                  HT, \label{eqn:rc0a} \\
  HT & = & \sigma T_{n}^4 e^{-\tau_{LW}} +
           \sigma T_{d}^4 \int_0^{\tau_{LW}}
           \left(\frac{\tau'}{\tau_{LW}}\right)^{4\beta}
           e^{-\tau'} d\tau', \label{eqn:rc0b} \\
  0 & = & \sigma T_{n}^4 -
          \sigma T_{d}^4 \int_0^{\tau_{LW}}
          \left(\frac{\tau'}{\tau_{LW}}\right)^{4\beta}
          e^{-(\tau_{LW}-\tau')} d\tau', \label{eqn:rc0c}
\end{eqnarray}
\end{subequations}
where $T_d$ is the dayside temperature, $T_n$ is the nightside
temperature, and $HT$ is the day-night heat transport. We express
the stellar flux in terms of the equilibrium temperature,
$L_*(1-\alpha_p)/2 = 2 \sigma T_{eq}^4$. Next, we combine the TOA
equations to eliminate $HT$ and use the nightside surface budget to write $T_n$
in terms of $T_d$. We find
\begin{subequations}
\begin{eqnarray}
  \sigma T_{d}^4 & = & \frac{ 2 \sigma T_{eq}^4 }{ 2 \int_0^{\tau_{LW}}
                       \left(\frac{\tau'}{\tau_{LW}}\right)^{4\beta}
                       e^{-\tau'} d\tau'
                       +e^{-\tau_{LW}} \left[ 1 +
                       \int_0^{\tau_{LW}} \left(\frac{\tau'}{\tau_{LW}}\right)^{4\beta}
                       e^{-(\tau_{LW}-\tau')} d\tau' \right]}, \label{eqn:rc1a}\\
  \sigma T_{n}^4 & = & \frac{ 2 \sigma T_{eq}^4 \times \int_0^{\tau_{LW}}
                       \left(\frac{\tau'}{\tau_{LW}}\right)^{4\beta}
                       e^{-(\tau_{LW}-\tau')} d\tau'}{ 2 \int_0^{\tau_{LW}}
                       \left(\frac{\tau'}{\tau_{LW}}\right)^{4\beta}
                       e^{-\tau'} d\tau'
                       +e^{-\tau_{LW}} \left[ 1 +
                       \int_0^{\tau_{LW}} \left(\frac{\tau'}{\tau_{LW}}\right)^{4\beta}
                       e^{-(\tau_{LW}-\tau')} d\tau' \right]}. \label{eqn:rc1b}
\end{eqnarray}
\label{eqn:rc1}
\end{subequations}
The first term in the denominator is the atmosphere's contribution to
the TOA flux, the second term is the TOA flux contribution from the dayside
and nightside surfaces. In practice we evaluate the definite integrals
in these expressions numerically, but they can also be expressed in terms of gamma functions
\citep{robinson2012}.

In the optically thick limit, these expressions reduce to the result
of \citet{pierrehumbert2011b}. For $\tau_{LW} \gg 1$ the
exponential terms $e^{-\tau_{LW}}$ become negligibly small. The
integrand in the upward flux decays exponentially at large $\tau'$,
which means we can approximate the upper limit as infinity and replace
the integral with a gamma function,
$\tau_{LW}^{-4\beta} \int_0^{\tau_{LW}} \tau'^{4\beta}e^{-\tau'}d\tau'
\approx \tau_{LW}^{-4\beta}
\int_0^{\infty}\tau'^{4\beta}e^{-\tau'}d\tau'=\tau_{LW}^{-4\beta}
\Gamma(1+4\beta)$.
Similarly, in the optically thick limit the downward flux at the
surface has to approach unity,
$\int_0^{\tau_{LW}}
(\tau'/\tau_{LW})^{4\beta}e^{-(\tau_{LW}-\tau')}d\tau' \approx 1$.
Combining these approximations we find
$T_n \approx T_d \approx T_{eq} \tau_{LW}^\beta
\Gamma(1+4\beta)^{-1/4}$,
which is the same as \citeauthor{pierrehumbert2011b}'s result
(Equation \ref{eqn:pierrehumbert1}). The dayside and nightside
temperatures become equal in this limit because the
atmosphere's downward longwave emission becomes large enough to
eliminate the temperature difference between the nightside surface and
the air directly above it (which in turn is equal to the dayside
surface temperature).

In the optically thin limit, our model differs slightly from the
result of \citet{wordsworth2015}. For $\tau_{LW} \ll 1$ we can
approximate all exponentials using Taylor series,
$e^{-\tau_{LW}}=1+\mathcal{O}(\tau_{LW})$. Retaining only the lowest
order in $\tau_{LW}$, we write the integrals in the upward and
downward fluxes both as
$\tau_{LW}^{-4\beta} \int_0^{\tau_{LW}}
\tau'^{4\beta}e^{\pm\tau'}d\tau' \approx
\tau_{LW}^{-4\beta}\int_0^{\tau_{LW}} \tau'^{4\beta} d\tau' =
\tau_{LW}/(1+4\beta)$.
The atmosphere's TOA upward and surface downward emission therefore
become equal, which is a well-known property of grey radiation in the
optically thin limit \citep{pierrehumbert2011b}. Again discarding
higher-order terms in $\tau_{LW}$, we find
$T_d \approx 2^{1/4} T_{eq}\times [1-3 \tau_{LW}/(4(1+4\beta))]$ and
$T_n \approx 2^{1/4} T_{eq} \tau_{LW}^{1/4} \times (1+4\beta)^{-1/4}$.
This nightside temperature has the same asymptotic limit but is
slightly warmer than Equation \ref{eqn:wordsworth1} from
\citet{wordsworth2015}. That is because we assume the atmosphere
remains fixed to an adiabat, whereas \citet{wordsworth2015} assumes an
atmosphere that is vertically isothermal. We will use our
radiative-convective-subsiding model (Sections
\ref{sec:radsub}-\ref{sec:transition}) to show that
\citeauthor{wordsworth2015}'s result is a limiting expression for
atmospheres that are very hot or thin, $t_{wave}/t_{rad} \gtrsim 1$, whereas our results in this
Section apply for atmospheres that are cold or thick,
$t_{wave}/t_{rad}\lesssim10^{-4}$.  Nevertheless, $\beta$ is always of
order unity so the difference between Equation \ref{eqn:wordsworth1} and our
result is small in the optically thin regime.

%%%%
\begin{figure}[h!]
%\figurenum{3}
\epsscale{0.9}
% \plotone{f3.eps}
\plotone{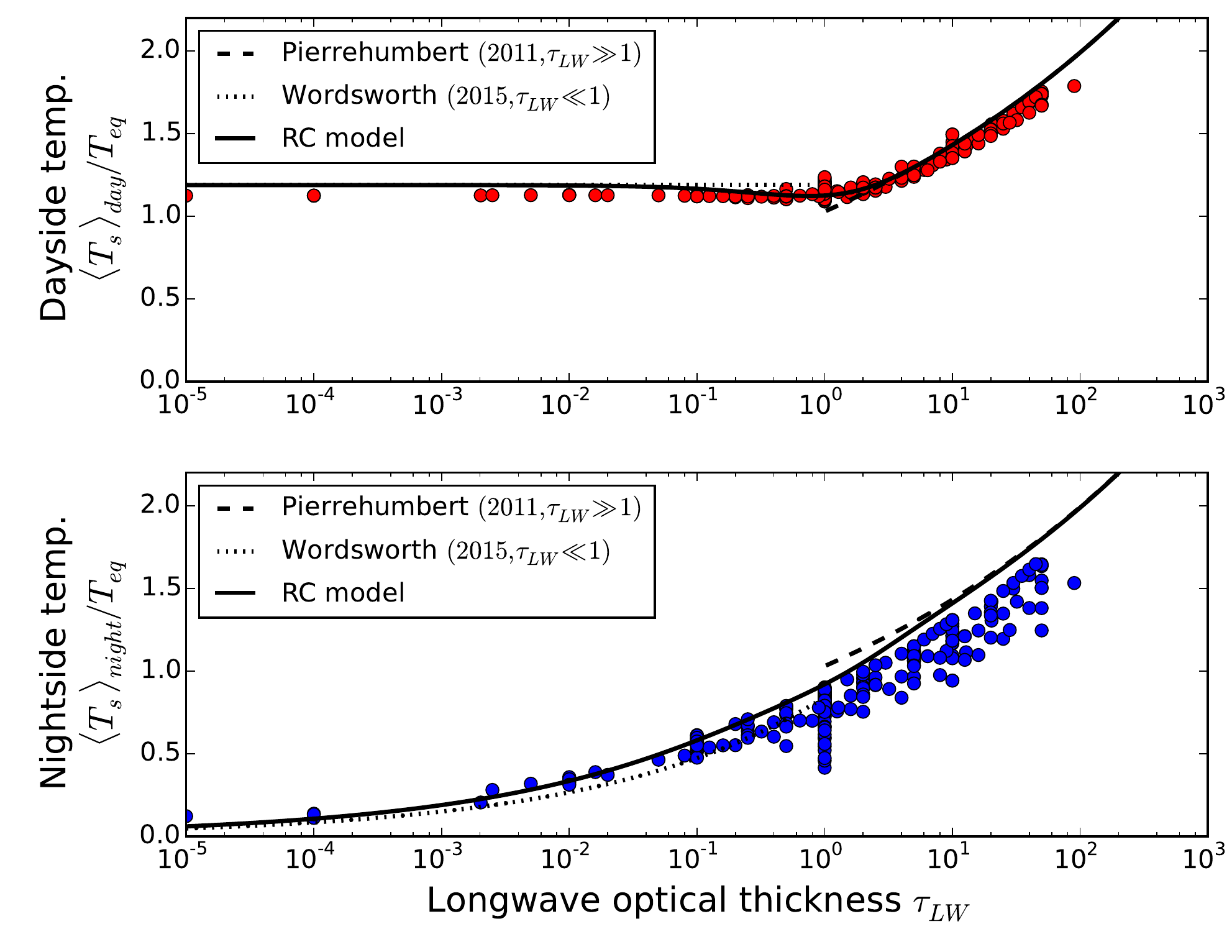}
  \caption{Our radiative-convective (RC) model captures the basic
    dependency of surface temperature on $\tau_{LW}$ and joins
    previous asymptotic limits. Top: Average dayside surface
    temperatures of many GCM simulations ($N=251$).
    Bottom: Average nightside surface temperatures. Dashed and dotted
    curves show previously-derived asymptotic scalings in the
    optically thick \citep[$\tau_{LW}\gg1$,][]{pierrehumbert2011b} and optically thin limits
    \citep[$\tau_{LW}\ll1$,][]{wordsworth2015}. The solid curve shows the RC model
    (Section \ref{sec:radconv}). While the RC model closely matches
    the GCM dayside temperatures, it does not account for the wide
    spread in nightside temperatures.  All shown simulations use $n=2$
    and $(R,c_p)=(R,c_p)_{N_2}$.
  \label{fig:Tsurf1}}
\end{figure}
%%%%

%%%%
\begin{figure}[h!]
%\figurenum{4}
\epsscale{0.9}
%\plotone{f4.eps}
\plotone{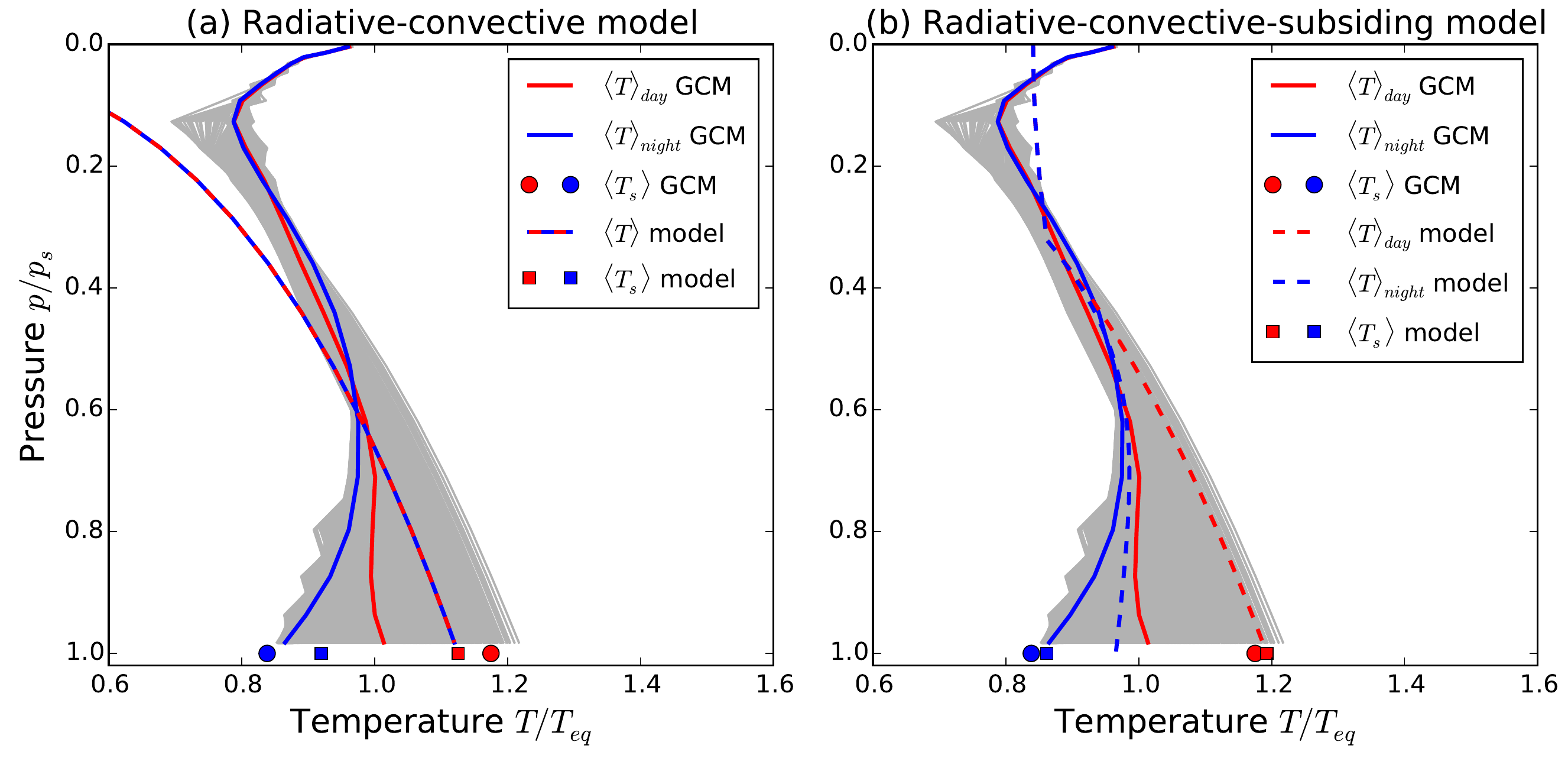}
  \caption{
    Temperature structure of a slowly rotating GCM simulation
    $(a^2/L_{Ro}^2=0.1)$, compared
    with the radiative-convective (RC, left) and the
    radiative-convective-subsiding model (RCS, right).
    Solid curves correspond to dayside (red) and
      nightside-averaged (blue) GCM temperature
    profiles, and GCM temperature profiles at each
    latitude and longitude (grey). Left: Although the RC model (mixed
    red-blue curve)
    qualitatively captures the temperature structure, it
    does not capture the nightside inversion and thus overpredicts the
    nightside surface temperature (compare blue square with blue circle).
    Right: the RCS model (dashed curves) accounts for imperfect
    day-night heat transport and qualitatively captures the nightside
    inversion structure. This leads to a better fit of nightside surface
    temperature than for the RC model. The
    planet's physical parameters are $T_{eq}=400\mathrm{K},a=a_\Earth,\Omega=2\pi/(50 \mathrm{d}),p_s=0.5
    \mathrm{bar},\tau_{LW}=1,(R,c_p)=(R,c_p)_{N_2}$, and $g=5
    \mathrm{m~s}^{-2}$.
  \label{fig:climDry50bc}}
\end{figure}
%%%%

Next, we compare the previous scalings and our radiative-convective
(RC) model with our GCM simulations. Figure \ref{fig:Tsurf1} shows
dayside (top) and nightside (bottom) average surface temperatures of
many simulations.
To represent all GCM results in a single figure, we normalize surface
temperatures using the equilibrium temperature $T_{eq}$
of each simulation. We only show simulations with $\beta = 1/7$,
i.e., $(R,c_p,n) = (R_{N_2},c_{p,N_2},2)$.  First, as we showed above,
the RC model tends towards the expressions of \citeauthor{pierrehumbert2011b}
and \citeauthor{wordsworth2015} in the optically thick and thin
regimes (compare solid line with dashed and dotted lines). While the
two approximate expressions diverge at $\tau_{LW}=1$, our model
provides a smooth fit in this region.  Second, the RC model captures
dayside surface temperatures very well, with deviations beween the RC
model and the GCM simulations smaller than $0.1 \times T_{eq}$. The RC
model systematically overpredicts dayside temperatures
because of its idealized geometry, which represents the entire dayside as a
single column. For example, the dayside-average temperature of an
airless planet in pure radiative equilibrium is
$4 \sqrt{2}/5 \times T_{eq}\approx 1.13 T_{eq}$, whereas the RC model
predicts $2^{1/4} \times T_{eq} \approx 1.19 T_{eq}$.
Third, the RC model captures the general trend of
nightside surface temperature with $\tau_{LW}$. However, Figure
\ref{fig:Tsurf1} also shows that nightside temperatures exhibit a much
wider spread than dayside temperatures, which is not captured by the RC model.

There are two reasons for the spread in nightside temperatures: first,
rapidly rotating atmospheres develop horizontally inhomogeneous
nightsides, and second, tidally locked atmospheres do not have an
adiabatic temperature structure on the nightside. We address rotation
in Section \ref{sec:rotation}, here we consider the effect of
temperature structure.  Figure \ref{fig:climDry50bc}a shows the
vertical temperature structure of a slowly rotating simulation. The
grey lines show the vertical temperature profiles at each horizontal
GCM grid point, which form a wide envelope. The hottest temperatures
at the right side of the envelope correspond to the substellar
point. These profiles are indeed adiabatic, which can be seen from the
fact that they are parallel to the temperature profile of the RC model
(dashed red-blue line). However, as the dayside and nightside averaged
profiles show, large parts of the atmosphere do \textit{not} follow an
adiabat (solid red and blue lines in Fig.~\ref{fig:climDry50bc}a).
The deviation arises because WTG breaks down inside the dayside
boundary layer (Fig.~\ref{fig:climDry50a}a). This allows the
atmosphere outside the boundary layer to decouple from regions of
convection, and develop a strongly non-adiabatic temperature profile.
In particular, Figure \ref{fig:climDry50bc} shows that the nightside
average (blue line) forms a strong inversion below $p/p_s \sim 0.6$,
which means the nightside is stably stratified and far from
radiative-convective equilibrium. Nightside inversions are a robust
feature of tidally locked atmospheres and have been found in a range
of simulations \citep[e.g.,][]{joshi1997,merlis2010,leconte2013}, but
are not captured by the RC model.  As a consequence the RC model
produces a warmer nightside atmosphere and therefore also a warmer
nightside surface than the GCM (compare blue square and blue circle in
Fig.~\ref{fig:climDry50bc}a). We present a model that captures the
nightside temperature structure in Section \ref{sec:radsub}. However,
to do so we have to account for atmospheric dynamics, which show up
via the parameters $t_{wave}/t_{rad}$ and $t_{wave}/t_{drag}$. To
address the dynamics we first have to develop a theory of large-scale
wind speeds and the atmospheric circulation, which we turn to in the
next section.

\section{A heat engine scaling for wind speeds}
\label{sec:winds}

Earth's atmosphere acts as a heat engine: it absorbs heat near the
surface at a high temperature and emits heat to space at a low
temperature, which allows the atmosphere to do work and balance
frictional dissipation \citep{peixoto1984}.  On Earth
the heat engine framework has been used to derive upper bounds on the strength of
tropical moist convection \citep{renno1996,emanuel1996} and
small-scale circulations such as hurricanes \citep{emanuel1986}.

%%%%
\begin{figure}[b!]
%\figurenum{5}
%\plotone{f5.pdf}
\begin{center}
 \includegraphics[width=.8\textwidth,natwidth=1012,natheight=642]{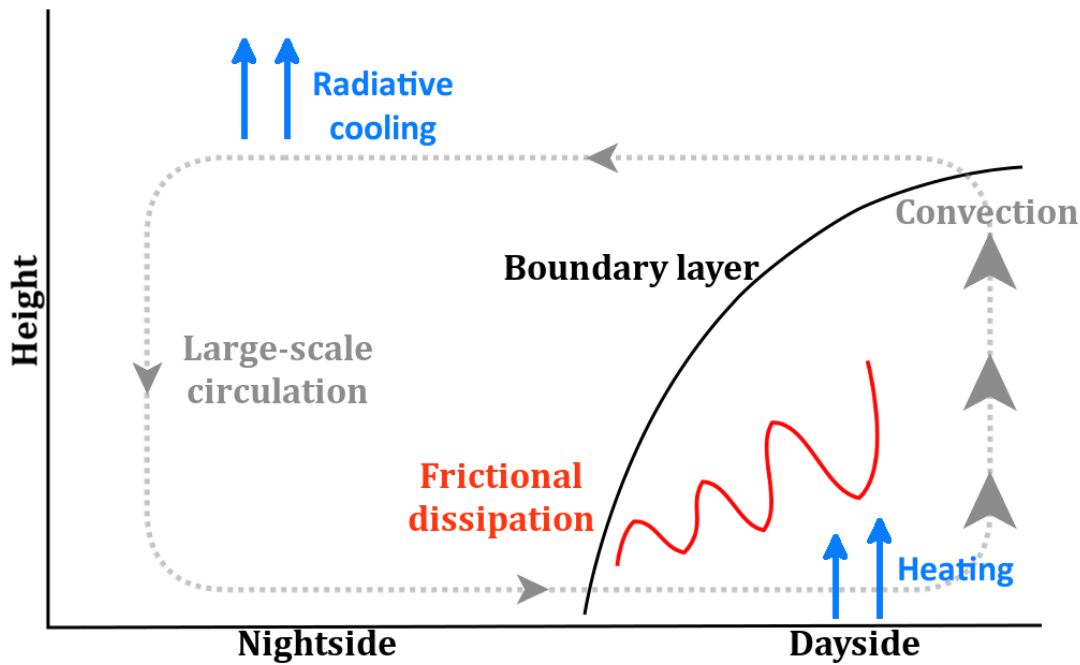}
\end{center}
  \caption{A diagram of the atmospheric heat engine. The heat engine
    is driven by dayside heating and cooling to space.  Frictional
    dissipation in the dayside boundary layer limits the strength of
    the resulting day-night atmospheric circulation.
  \label{fig:diagHEngine}}
\end{figure}
%%%%
In this section we idealize the atmospheric circulation of a tidally
locked planet as a single overturning cell between the substellar and
antistellar point. We model the circulation as an ideal heat engine to
place an upper bound on its circulation strength. The ideal heat engine is an upper
bound because additional physical processes, such as diffusion, lead
to irreversible production of entropy and decrease the efficiency of a
heat engine below its ideal limit \citep{pauluis2002a}.  As shown in
Figure \ref{fig:diagHEngine}, the atmosphere absorbs heat near the
dayside surface at a hot temperature and emits it to space at a cold
temperature. These temperatures are defined in terms of
entropy-weighted averages over which the atmosphere absorbs and gives
off heat \citep{emanuel1989,pauluis2002a}.
Here we idealize the dayside as a single column that follows an
adiabat. Entropy is therefore vertically constant on the dayside,
which means the temperature at which the atmosphere absorbs heat is equal to the
dayside surface temperature, $T_{d}$. We approximate the cold
temperature as the planet's effective emission temperature to space,
i.e., its equilibrium temperature $T_{eq}$.  The parcel does work
against friction in the boundary layer which is given by
$W = C_D \rho_s U_s^3$ \citep{bister1998}. Here $W$ is the work,
$\rho_s$ is the surface density, and $U_s$ is a surface wind speed,
which we take to be the dayside-average surface wind
(Fig.~\ref{fig:diagHEngine}). Using Carnot's theorem,
\begin{eqnarray}
  W & = & \eta Q_{in},
\end{eqnarray}
where $\eta = (T_{d}-T_{eq})/T_{d}$ is the atmosphere's thermodynamic
efficiency, and $Q_{in}=2\sigma T_{eq}^4 \times (1-e^{-\tau_{LW}})$ is
the amount of energy that is available to drive atmospheric motion. We
note that the dayside-averaged incoming stellar flux is equal to
$2\sigma T_{eq}^4$, but we additionally account for the fact that only
a fraction $1-e^{-\tau_{LW}}$ of stellar energy is available to the
atmosphere, while the remainder is immediately re-radiated from the
surface to space.

We find the following upper bound on the dayside average surface wind speed,
\begin{eqnarray}
  U_s & = & \left[ \frac{T_{d} - T_{eq}}{T_{d}}
                   \times (1-e^{-\tau_{LW}})
                   \frac{2\sigma T_{eq}^4}{C_D \rho_s} \right]^{1/3} \nonumber \\
                 & = & \left[ \frac{T_{d} - T_{eq}}{T_{d}} \times (1-e^{-\tau_{LW}})
                   \frac{2 R T_{d} \sigma T_{eq}^4}{C_D p_s} 
                   \right]^{1/3} \nonumber \\
                 & = & \left[ (T_{d} - T_{eq}) \times (1-e^{-\tau_{LW}})
                   \frac{ 2 R \sigma T_{eq}^4}{C_D p_s}     \right]^{1/3},
\label{eqn:he1}
\end{eqnarray}
where we used the ideal gas law to substitute for $\rho_s$ in the
second step.
The only unknown in this equation is the dayside surface temperature
$T_{d}$. As we saw in Section \ref{sec:radconv}, $T_{d}$ was already
well constrained by the radiative-convective model (Fig.~\ref{fig:Tsurf1}). In this section we therefore close
the model using $T_{d}$ from Equation \ref{eqn:rc1a} (but note that
we will self-consistently solve for $T_{d}$ in Section \ref{sec:radsub}).

%%%%
\begin{figure}[htb!]
%\figurenum{6}
%  \plotone{f6.eps}
\plotone{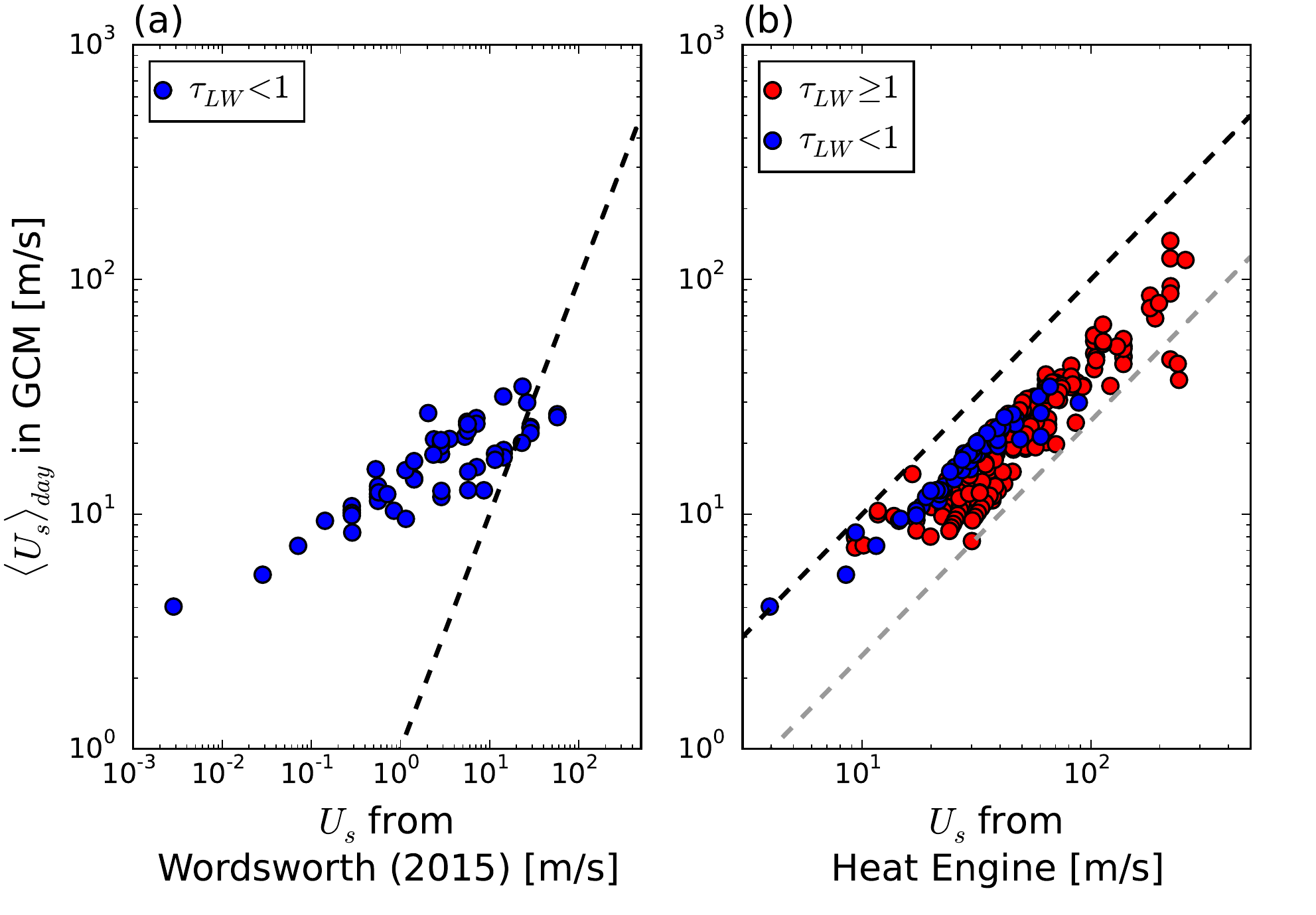}
  \caption{Two different surface wind speed scalings compared with
    many GCM results ($N=271$).
    Left: scaling for dayside surface wind speed from
    \citet{wordsworth2015}, which was derived assuming
    optically thin atmospheres ($\tau_{LW}<1$). Right: our scaling for average surface wind
    speed for an ideal heat engine (Eqns.~\ref{eqn:he1} and
    \ref{eqn:rc1a}). The GCM simulations are less efficient than ideal
    heat engines and therefore have smaller surface wind speeds. The
    grey dashed line corresponds to an inefficiency factor of $1/4$.
  \label{fig:Usurf1}}
\end{figure}
%%%%

Figure \ref{fig:Usurf1} compares dayside averaged surface wind speeds
$\langle U_s\rangle$ with a numerical wind speed scaling from
\citet[][see Appendix \ref{sec:appendix2}]{wordsworth2015} and our
analytical heat engine scaling. We note that
\citeauthor{wordsworth2015} considered the optically thin limit,
whereas our results are valid for arbitrary
$\tau_{LW}$. \citeauthor{wordsworth2015} derived a scaling by assuming
weak temperature gradients hold globally. In a
weak-temperature-gradient (WTG) atmosphere, radiative cooling leads to
subsidence, which \citeauthor{wordsworth2015} assumed in turn drives
the large-scale circulation. Figure \ref{fig:Usurf1} shows that the
GCM wind speeds span two orders of magnitude, from $3$ m s$^{-1}$ up
to about $300$ m s$^{-1}$. The \citet{wordsworth2015} scaling seems to
match these wind speeds at $\mathcal{O}(1)$ m s$^{-1}$. However, it
predicts a strong decrease, down to less than 10$^{-2}$ m s$^{-1}$,
which is several orders of magnitude smaller than the GCM results
(Fig.~\ref{fig:Usurf1}a). The mismatch arises because
\citeauthor{wordsworth2015}'s global WTG scaling assumes winds are
purely driven by radiative cooling, $U_s \propto \tau_{LW}$ (Appendix
\ref{sec:appendix2}), so $U_s$ should rapidly vanish in the optically
thin limit.
Instead, Figure \ref{fig:climDry50a} shows that WTG balance breaks down in regions that are
strongly convecting. The convecting regions in turn govern the
return flow from the nightside to the dayside, which means that the
effect of friction on the large-scale circulation cannot be neglected.
Our heat engine scaling includes this effect and predicts very
different dynamics. For example, in the
optically thin limit the dayside temperature is approximately constant
and $1-e^{\tau_{LW}}\approx \tau_{LW}$, so $U_s\propto
\tau_{LW}^{1/3}$ (Equation \ref{eqn:he1}).
Figure \ref{fig:Usurf1}b supports our theory. The slope predicted by
the heat engine provides an excellent fit to the GCM
results. Moreover, we expect the heat engine to provide an upper bound
on surface wind speeds. Our expectation is confirmed by the GCM
simulations, which fall below the dashed black line in Figure
\ref{fig:Usurf1}b.
In addition, the overestimate of $\langle U_s\rangle$ is small and generally
amounts to less than a factor of 4 (grey dashed line in Figure
\ref{fig:Usurf1}b), with most simulations falling about a factor of 2
below the ideal limit.

Next, we use the surface wind speed scaling to place an upper bound on
the strength of the day-night circulation. Of particular interest to us is the large-scale
vertical motion on the nightside, which we will
show governs the day-night heat transport and is critically important for the temperature
structure on the nightside (Section \ref{sec:radsub}).
We express all vertical motions using pressure
coordinates, that is, using the pressure velocity $\omega \equiv
D p/Dt$ where $\omega>0$ means sinking motions.
We take the surface wind speed $U_s$ to be the
characteristic horizontal velocity within the boundary layer. We relate
the horizontal velocity in the boundary layer to the pressure velocity
near the substellar point using mass conservation (Equation \ref{eqn:consmass}),
\begin{eqnarray}
  \frac{\omega_{up}}{p_s} \sim \frac{U_s}{a}.
\end{eqnarray}
Figure \ref{fig:climDry50a}b supports this scaling, and $\omega/\omega_{up}$
near the substellar point is of order unity.
However, Figure \ref{fig:climDry50a}b also shows that there is a large
asymmetry between rising and sinking motions.  While air rises rapidly
near the region of strongest convection at the substellar point, it
sinks slowly over a large area outside the boundary layer. Figure
\ref{fig:climDry50a}c quantifies the asymmetry using
$A_{up}/A_{down}$, the fraction of the atmosphere in which air rises
versus sinks\footnote{Because the uppermost layers of the atmosphere
  show both weakly rising and falling motions we define $A_{up}$ as
  the area with ``significant'' upward motion where
  $\omega \leq 0.01 \times \mathrm{min}(\omega)$.}. In the shown simulation
rising air never covers more than 20\% of the atmosphere,
while its vertically averaged value is about 10\% (dot in
Fig.~\ref{fig:climDry50a}c).  The asymmetry in vertical motions arises
from the geometric asymmetry of the incoming stellar flux, and is distinct
from the asymmetry of rising and sinking motions in Earth's tropics
which is caused by the condensation of water during convection.
Because upward and downward mass fluxes
have to balance across a horizontal slice of atmosphere,
\begin{eqnarray}
  \rho A_{up} \omega_{up} = \rho A_{down} \omega_{down},
\end{eqnarray}
where $\rho$ is the density of an air parcel, we can relate the
pressure velocity on the nightside to the dayside surface wind,
\begin{eqnarray}
  \omega_{down} = \frac{A_{up}}{A_{down}} \frac{p_s}{a} \times U_s.
\label{eqn:he2}
\end{eqnarray}
Equation \ref{eqn:he2} explains how tidally locked planets sustain
weak downward motions despite very large horizontal wind speeds. The
time for a parcel of air to be advected horizontally is
$t_{adv} = a/U_s$ whereas the time for a parcel to subside (that is,
be advected vertically) is
$t_{sub} = p_s/\omega_{down} = A_{down}/A_{up} \times t_{adv}$. For
$A_{down}/A_{up}\sim10$ it takes a parcel of air ten times longer to
sink back to the surface on the nightside than to be advected from the
nightside to the dayside. The same comparison also explains why day-night temperature
gradients of tidally locked planets are not set by the advective
timescale and instead depend on the ratio of subsidence and radiative
timescales (Section \ref{sec:transition}).

%%%
\begin{figure}[t!]
%\figurenum{7}
\epsscale{0.75}
%  \plotone{f7.eps}
\plotone{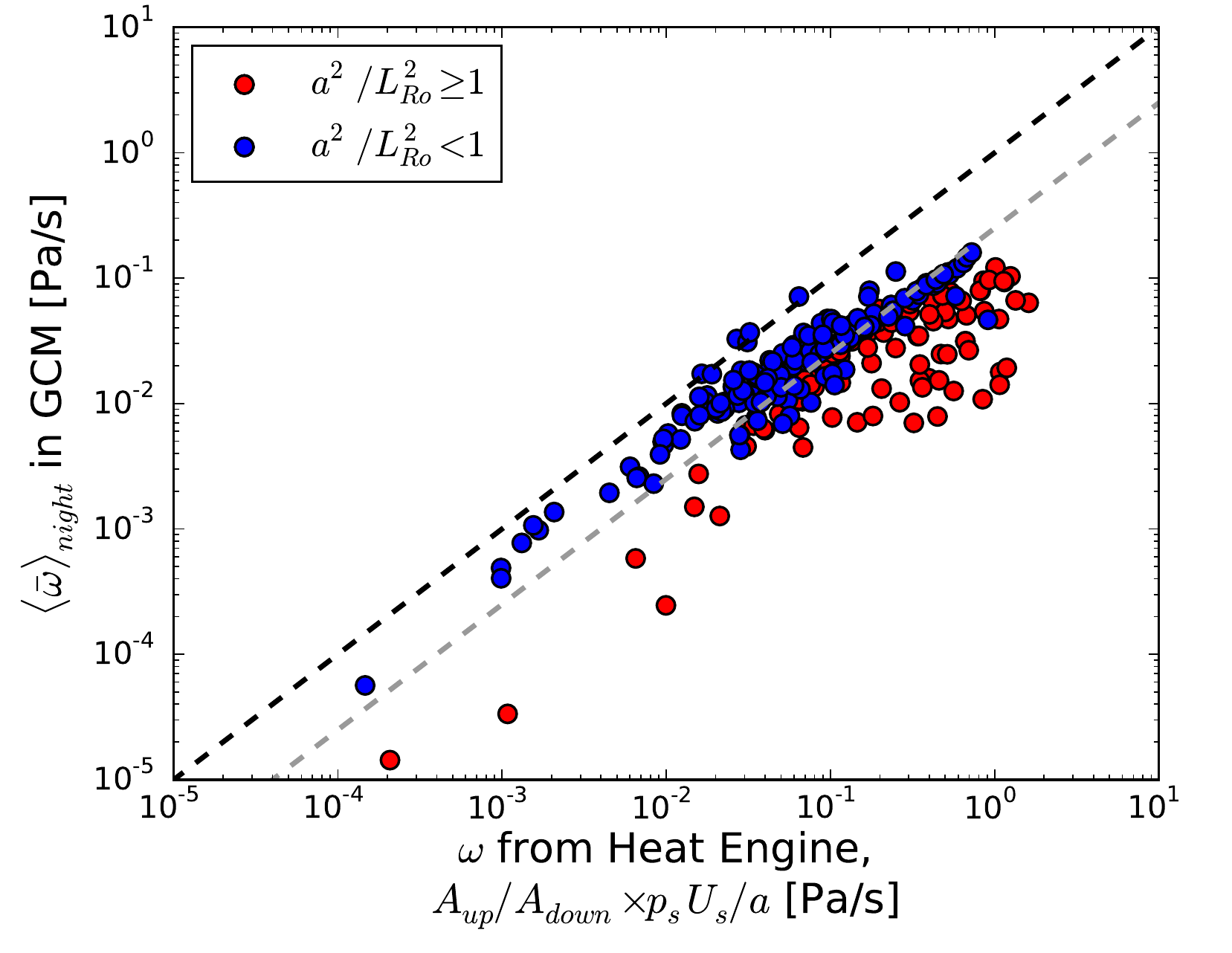}
  \caption{The heat engine scaling provides a strong constraint on the
  day-night atmospheric circulation. Shown is the
  vertical velocity in pressure coordinates predicted by the heat engine scaling
  (x-axis), compared with the average nightside pressure velocity in
  the GCM (y-axis).
  Rapidly rotating atmospheres, $a^2/L_{Ro}^2\geq1$, develop
  inhomogeneous nightsides and can locally sustain smaller pressure
  velocities (Section \ref{sec:rotation}).
  \label{fig:Omega1}}
\end{figure}
%%%%

Next, Figure \ref{fig:Omega1} compares the pressure velocity
$\omega_{down}$ from Equation \ref{eqn:he2} with the mass-weighted
vertically and horizontally averaged pressure velocities,
$\langle \bar{\omega} \rangle_{night}$, from GCM
simulations. In the comparison we use Equation \ref{eqn:he1} to
predict $U_s$ but still diagnose $A_{up}/A_{down}$ directly from GCM
output.
First, because the heat engine provides an upper limit on $U_s$ it
also provides an upper limit on $\omega_{down}$. The GCM simulations
indeed fall almost entirely below the dashed black line in Figure
\ref{fig:Usurf1}b. We note that in deriving Equation \ref{eqn:he1} we
neglected some factors that we expect to be small (e.g., geometric
factors), but which explain why some GCM simulations
slightly exceed the value predicted by the scaling.  Second, we find
that relatively slowly rotating atmospheres (blue
dots) closely follow the heat engine scaling and most of them deviate
less than a factor of 4 from it (grey dashed line). Third, rapidly
rotating atmospheres (red dots) still follow the
scaling qualitatively but $\langle \bar{\omega} \rangle_{night}$ is
smaller than in slowly rotating atmospheres. The larger deviation
arises because rapidly rotating atmospheres develop inhomogeneous nightsides (Section
\ref{sec:rotation}). In the extra-tropics the flow then becomes
geostrophic which in turn suppresses vertical motions by
$\mathcal{O}(Ro)\ll1$, where $Ro$ is the Rossby number \citep{showman2010}.

We conclude that atmospheres of dry tidally locked planets are
dominated by dayside boundary layer friction. The heat engine
framework successfully constrains the amount of dissipation and
surface wind speeds within the boundary layer. Combined with the
areal asymmetry between rising and sinking motions, we find an upper
bound on the nightside vertical velocity. Our result is
distinct from previous scalings that have been proposed for
exoplanets. We will use our result in the next section to constrain
the thermal structure of the nightside.

\section{A two-column radiative-convective-subsiding model}
\label{sec:radsub}

%%%%
\begin{figure}[htb!]
%\figurenum{8}
%\plotone{f8.pdf}
\begin{center}
\includegraphics[width=.8\textwidth,natwidth=975,natheight=640]{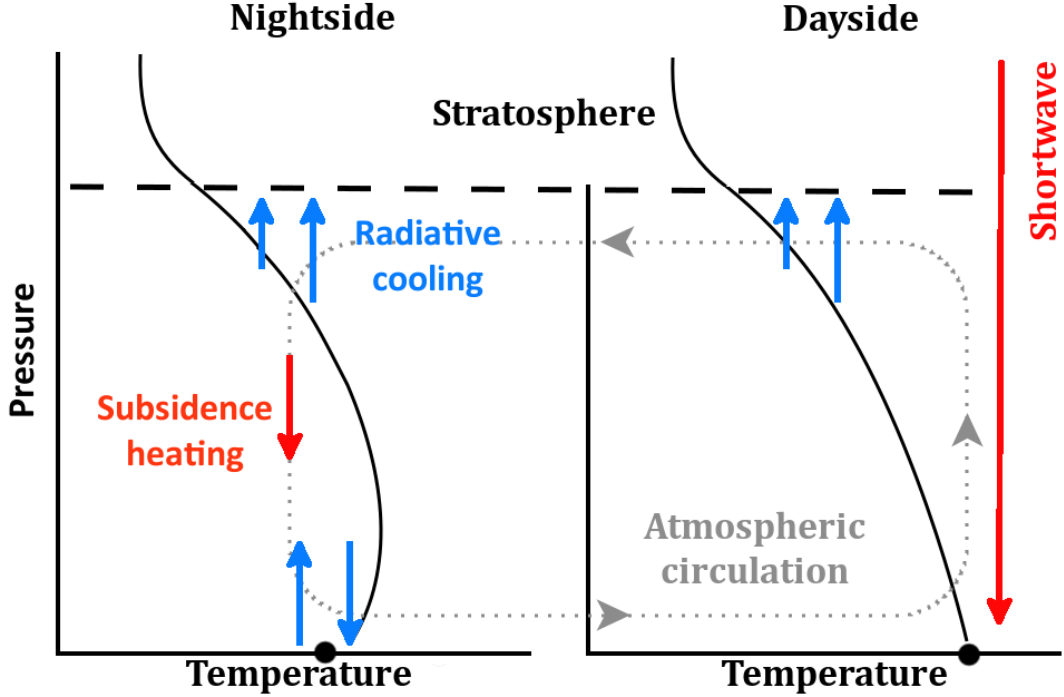}
\end{center}
\caption{A diagram of the two-column radiative-convective-subsiding model. We assume
    convection sets an adiabatic temperature profile on the dayside,
    and a balance between radiative cooling and subsidence heating sets the
    temperature profile on the nightside. In addition, both columns
    are capped by a horizontally uniform and purely radiative
    stratosphere. The day-night circulation and the rate of subsidence
    are governed by the atmospheric heat engine.
  \label{fig:diagRadSub}}
\end{figure}
%%%%

As we showed in Figures \ref{fig:Tsurf1} and \ref{fig:climDry50bc}, to
understand nightside surface temperatures of tidally locked planets we
need to account for an imperfect day-night heat transport and to
better constrain the nightside atmospheric temperature structure. In
this section we develop a two-column model that does so. We again
divide the atmosphere into two dayside and nightside columns, shown in
Figure \ref{fig:diagRadSub}.  As in Section \ref{sec:radconv} the
dayside column is strongly convecting, but we allow the nightside
temperature profile to deviate from an adiabat. Both columns are
capped by a stratosphere, that is, a layer in pure radiative
equilibrium.

As in our radiative-convective model, convection sets an adiabatic
temperature profile on the
dayside. The dayside temperature profile is therefore
\begin{eqnarray}
  T & = & T_{d} \left( \frac{\tau}{\tau_{LW}} \right)^{\beta}.
\label{eqn:adiabat2}
\end{eqnarray}

The nightside is in weak-temperature-gradient (WTG)
balance. WTG balance follows from the thermodynamic
equation (Equation \ref{eqn:consenergy}),
\begin{eqnarray}
\frac{\partial T}{\partial t} + \mathbf{u} \cdot \nabla T + \omega \frac{\partial T}{\partial p} &=&
\frac{R T \omega}{c_p p} + \frac{g}{c_p}\frac{\partial F }{\partial p}
+ \frac{g}{c_p}\frac{\partial \mathcal{D}}{\partial p},
\end{eqnarray}
where $\mathbf{u}$ is the horizontal velocity, $\omega$ is the
pressure velocity ($\omega > 0$ for subsiding air), $F$ is the net
radiative flux (the sum of upward and downward longwave
fluxes, $F=F^{\uparrow}-F^{\downarrow}$), and $D$ is the energy flux
due to diffusion. The left side of the thermodynamic equation
represents advection, the first term on the right is
heating/cooling due to compression/expansion as air parcels move
vertically, the second term on the right is radiative heating/cooling,
and the third term represents the effect of small-scale convection
inside the boundary layer.
In equilibrium $\partial T/\partial
t=0$, and $\mathcal{D}$ is negligible on the nightside
  because the nightside is stably stratified. As long as
horizontal temperature gradients are small on the nightside the
thermodynamic equation then reduces to WTG balance
\begin{eqnarray}
  \omega \left( \frac{\partial T}{\partial p} - \frac{R T}{c_pp}
  \right)
  &\approx& \frac{g}{c_p}\frac{\partial F}{\partial p}.
            \label{eqn:x1}
\end{eqnarray}
Equation \ref{eqn:x1} entails that radiative cooling is accompanied by
subsidence as follows. In a cooling layer the net radiative flux decreases towards the
surface, $\partial F/\partial p < 0$. The lapse
rate has to be smaller than, or equal to, the adiabatic lapse rate because the
atmosphere would otherwise start convecting, $\partial T/\partial p \leq R T/(c_p p)$.
It follows that $\omega > 0$.

In Earth's tropics the vertical temperature structure,
$\partial T/\partial p$, and radiative fluxes,
$\partial F/\partial p$, are set by small regions of moist convection,
in which case WTG can be used to predict the large-scale $\omega$
\citep{sobel2001}. In this section we pursue the opposite approach:
because $\omega$ is set by the day-night circulation which, in turn, is
limited by friction in the dayside boundary layer (Section
\ref{sec:winds}), we will use WTG balance to solve for $T$ and
$F$. For simplicity we replace $\omega$ with its vertical average
$\bar{\omega}$. Because we assume that horizontal variations are small,
we also replace all partial derivatives with normal derivatives.
We rewrite WTG balance in optical depth
coordinates and combine it with the Schwarzschild equation for the
radiative flux $F$ (Equation \ref{eqn:twostream}),
\begin{subequations}
\begin{eqnarray}
 \frac{c_p \bar{\omega}}{g} \left( \frac{d T}{d \tau} - \frac{\beta
  T}{\tau} \right)&=& \frac{d F }{d \tau},\\
  \frac{d^2 F}{d \tau^2} - F &=& -2 \frac{d(\sigma T^4)}{d \tau}.
\end{eqnarray}
\label{eqn:rcs1}
\end{subequations}
Given boundary conditions, these equations can be solved for $T$ and
$F$. The left side of Equation \ref{eqn:rcs1}a represents the vertical
energy flux due to subsidence (in W m$^{-2}$). In the WTG regime
subsidence is how the atmosphere transports heat between dayside and
nightside. Atmospheres with strong subsidence (large $\bar{\omega}$)
will tend to have nightsides that are close to an adiabat, while
atmospheres with very weak subsidence will tend to approach pure
radiative equilibrium on their nightsides (i.e.,
$d F/d\tau \approx0$).

To solve for $T$ and $F$ on the nightside we need to specify an upper
boundary condition. A natural choice is the tropopause, $\tau_0$, up
to which convection rises on the dayside. Above $\tau_0$ the
atmosphere is in pure radiative equilibrium, $d F/d\tau =0$. We
assume the stratosphere is horizontally uniform, which means it has
the same temperature structure as in \citet{pierrehumbert2011b},
\begin{eqnarray}
T_{strat} & = & T_{eq} \left(\frac{1+\tau}{2}\right)^{1/4}.
\label{eqn:rcs2}
\end{eqnarray}

We can now specify the boundary conditions for the nightside
atmosphere and Equations \ref{eqn:rcs1}. 
Because WTG balance is a first-order equation and the radiative
equation is a second-order equation we require three conditions,
\begin{subequations}
\begin{eqnarray}
  T(\tau_0) &=& T_{strat}(\tau_0),\\
  d F(\tau_0)/d\tau  &=& 0, \\
  F(\tau_{LW}) &=& 0.
\end{eqnarray}
\label{eqn:rcs4}
\end{subequations}
The first equation is temperature continuity at the tropopause. The
second is the stratospheric energy budget, that is, pure radiative
equilibrium.  The third condition is the nightside surface energy
budget. Because the nightside surface is in radiative equilibrium with
the overlying atmosphere,
$F^{\uparrow}(\tau_{LW})=F^{\downarrow}(\tau_{LW})$, the net radiative
flux $F=F^{\uparrow}-F^{\downarrow}$ has to vanish at the surface.
The only unknown in these boundary conditions is the tropopause height
$\tau_0$.

The tropopause height $\tau_0$ is in turn governed by convection on
the dayside. On the dayside, the
convective temperature profile (Equation \ref{eqn:adiabat2}) has to match the
stratospheric temperature profile (Equation \ref{eqn:rcs2}) at
$\tau_0$, so
\begin{eqnarray}
  T_{d} \left(\frac{\tau_0}{\tau_{LW}}\right)^{\beta} &=& T_{strat}
                                                          (\tau_0), \nonumber \\
  T_{d} \left(\frac{\tau_0}{\tau_{LW}}\right)^{\beta} &=& T_{eq} \left(  \frac{1+\tau_0}{2} \right)^{1/4}.
\label{eqn:rcs3}
\end{eqnarray}

Finally, we use the top-of-atmosphere (TOA) energy budget to constrain
$T_d$. The global TOA energy budget is 
\begin{eqnarray}
  2\sigma T_{eq}^4 & = & F(0)_{day}+F(0)_{night}.
\end{eqnarray}
The left side is the incoming solar radiation and the right side
is the dayside and nightside outgoing longwave radiation (OLR).
To specify these fluxes we note that the stratosphere is in radiative
equilibrium, $dF/d\tau=0$, so the OLR has to match the net flux at the
tropopause, $F(0) = F(\tau_0)$. 
The net radiative flux at the dayside tropopause is
\begin{eqnarray}
  F(\tau_0)_{day} & = & \sigma T_{d}^4 e^{-(\tau_{LW}-\tau_0)} +
                         \sigma T_{d}^4 \int_{\tau_0}^{\tau_{LW}}
                         \left(\frac{\tau'}{\tau_{LW}}\right)^{4\beta}
                         e^{-(\tau'-\tau_0)} d\tau' -\sigma T_{eq}^4 \frac{\tau_0}{2}.
\end{eqnarray}
The first two terms are the upwelling flux at the dayside tropopause
(from the surface and atmosphere respectively), and the third 
term is downward flux from the stratosphere \citep{robinson2012}. The
global TOA energy budget therefore is,
\begin{eqnarray}
  2\sigma T_{eq}^4 & = & \sigma T_{d}^4 e^{-(\tau_{LW}-\tau_0)} +
                         \sigma T_{d}^4 \int_{\tau_0}^{\tau_{LW}}
                         \left(\frac{\tau'}{\tau_{LW}}\right)^{4\beta}
                         e^{-(\tau'-\tau_0)} d\tau' -\sigma T_{eq}^4 \frac{\tau_0}{2}+ F(\tau_0).
\label{eqn:rcs5}
\end{eqnarray}
Equations \ref{eqn:rcs4}-\ref{eqn:rcs5} determine the
tropopause height $\tau_0$, the dayside surface temperature $T_d$, and
the nightside OLR $F(\tau_0)$.

Finally, we constrain the pressure velocity $\bar{\omega}$ on the
nightside. We showed in Figure \ref{fig:Omega1} that the heat
engine scaling allows us to place an upper bound on $\bar{\omega}$ once we account
for the fact that atmospheres are imperfect heat engines and once we
know the relative fraction of rising versus subsiding motions,
$A_{up}/A_{down}$ (we consider rotation in Section \ref{sec:rotation}).
In this section we incorporate these effects via
\begin{eqnarray}
  \omega_{down} = \chi \times \frac{p_s U_s}{a}
\label{eqn:rcs6}
\end{eqnarray}
where $\chi$ captures the inefficiency of the heat engine as well as
the smallness of $A_{up}/A_{down}$. We again use Equation
\ref{eqn:he1} to compute $U_s$, but now we self-consistently solve for
the dayside temperature $T_d$.  To constrain $\chi$ we note that the
asymmetry between rising and sinking motions is set by the tidally
locked geometry and hence should not vary much between different
simulations. We use $A_{up}/A_{down} \approx 0.1$ from
Figure \ref{fig:climDry50a}c as a representative value. Similarly, for
slowly rotating atmospheres we found that $\bar{\omega}$ falls between
the value predicted by the heat engine and about a factor of four less
(Fig.~\ref{fig:Omega1}), so we choose a representative inefficiency of
$1/2$. Combining these two, we find $\chi = 1/20$. Because rapidly
rotating atmospheres tend to have weaker nightside subsidence
(Figure \ref{fig:Omega1}), our choice of $\chi$ is an upper bound for
$\omega_{down}$ and will overestimate the day-night heat
transport on rapidly rotating planets. We emphasize that $\chi$ is
the only tunable parameter in our model and is fixed to a single
value. We do not change $\chi$ when we compare the
radiative-convective-subsiding model with different GCM simulations.

We numerically solve the model to find the nightside
  temperature $T$ and radiative flux $F$, the dayside surface
  temperature $T_d$, the nightside surface temperature $T_n$ and the
  tropopause height $\tau_0$. The boundary conditions for $T$ and $F$
  are specified at the tropopause and at the surface, so we use a
  shooting method (Appendix \ref{sec:appendix3}). We note that the
  Schwarzschild equation (Equation \ref{eqn:rcs1}b) becomes difficult
  to solve accurately in the optically thick limit because the
  radiative boundary conditions at the tropopause and surface decouple
  at large $\tau_{LW}$ (Equations \ref{eqn:rcs4}b,c). Nevertheless, the
  underlying physics do not change qualitatively once
  $\tau_{LW}\gg1$. We therefore avoid these issues by limiting our
  numerical solver to atmospheres with $\tau_{LW} \leq 15$.

Figure \ref{fig:climDry50bc}b compares the
radiative-convective-subsiding (RCS) model with the same slowly
rotating GCM simulation as in Figure \ref{fig:climDry50bc}a. The RCS model produces an adiabatic temperature profile on
the dayside and an inversion on the nightside. Compared to the
radiative-convective model (RC, Fig.~\ref{fig:climDry50bc}a), the RCS
model produces a colder nightside and a warmer dayside because it does
not assume that the day-night heat transport is necessarily highly effective.
The predicted temperatures match the GCM significantly better,
particularly on the nightside.
The RCS model also places the tropopause at $p/p_s \sim 0.3$, whereas the GCM tropopause
is higher up, at $p/p_s \sim 0.1$. The high tropopause in the GCM arises because
it is set by the deepest convection and hottest temperatures
near the substellar point instead of the average dayside temperature
(Fig.~\ref{fig:climDry50bc}b), which the RCS model does not account for.
Finally, the inversion
structure in the RCS model is somewhat skewed compared with the GCM,
and the inversion occurs higher up in the atmosphere
(Fig.~\ref{fig:climDry50bc}b). The raised inversion is likely due to our assumption of
a vertically constant value of $\omega$.  Nevertheless, given the
simplicity of the RCS model, we consider the fit between the RCS model
and the GCM highly encouraging. We emphasize that the RCS model is
obtained via a simple numerical solution, and is conceptually much
simpler (and computationally much cheaper) than the full GCM.

%%%
\begin{figure}[htb!]
%\figurenum{9}
%  \plotone{f9.eps}
\plotone{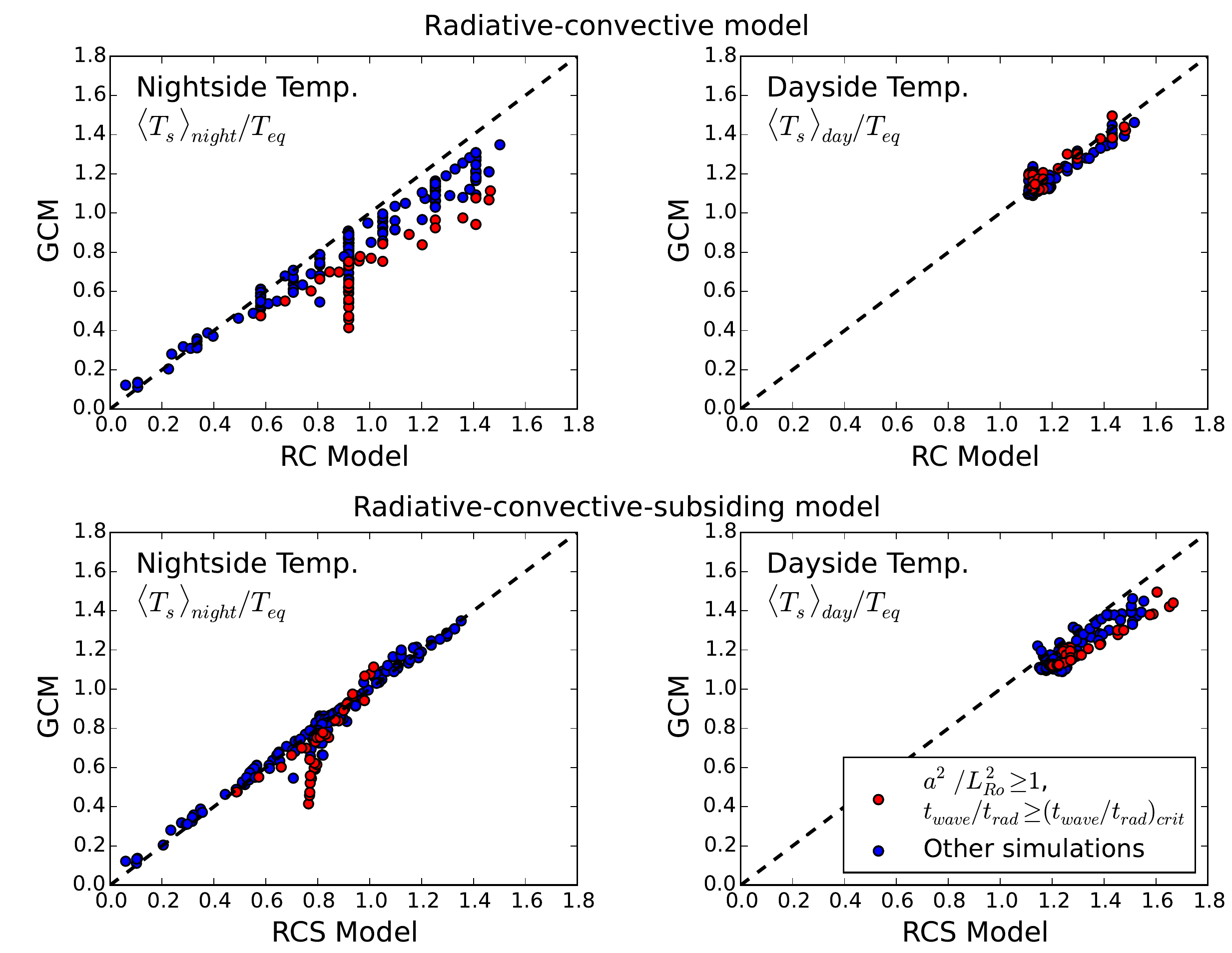}
  \caption{Comparison of surface temperatures predicted by the radiative-convective (RC) and
    radiative-convective-subsiding (RCS) models with many GCM
    simulations ($N=241$).
    Red dots represent simulations that are both rapidly rotating and
    have hot/thin atmospheres ($a^2/L_{Ro}^2 \geq 1$ and $t_{wave}/t_{rad}$ exceeds
    threshold from Equation \ref{eqn:transition3}), blue dots show all other simulations.
    Top left: average nightside temperature, RC
    model vs.~GCM.
    Top right: average dayside temperature, RC
    model vs.~GCM.
    Bottom left: average nightside temperature, RCS
    model vs.~GCM.
    Bottom right: average dayside temperature, RCS
    model vs.~GCM.
    The RCS model captures nightside temperatures much better than the
    RC model. The RCS model breaks down only for atmosphere that are
    both rapidly rotating and hot/thin (red dots; see Section
    \ref{sec:rotation}).
  \label{fig:Tsurf2}}
\end{figure}
%%%

Next, Figure \ref{fig:Tsurf2} compares the RC and RCS models with many
GCM simulations. The top row compares the radiative-convective model (RC)
with the GCM, the bottom row does the same for the
radiative-convective-subsiding (RCS) model.
We note that in rapidly rotating atmospheres ($a^2/L_{Ro}^2\geq 1$)
WTG balance does not hold at higher latitudes, and both models
should break down.
However, WTG balance actually provides a good
approximation of the nightside structure even at rapid rotation
provided the atmosphere is not too hot or thin
($t_{wave}/t_{rad}\gtrsim 5\times10^{-2}$). We explain this threshold
in Sections \ref{sec:transition} and \ref{sec:rotation}, here we
simply mark simulations for which the RCS model could break down in red and
all other simulations in blue.
First, as we already explained for Figure \ref{fig:climDry50bc}b, the
RCS model generally predicts warmer daysides than the RC model, which
already overestimates dayside temperatures slightly (see right panels
in Fig.~\ref{fig:Tsurf2}). To quantify the goodness of fit between the
GCM and our models, we compute $r^2$ values for the simulations marked
in blue. For dayside temperatures we find $r^2=0.82$ with the RC
model, and $r^2=0.23$ with the RCS model. These values underline that
the RC model already captures the basic structure of the
dayside. Improving the fit even further would require addressing the
spatial inhomogeneity on the dayside (Fig.~\ref{fig:climDry50a}),
whereas the reduced heat transport in the RCS model actually
worsens its dayside fit.
Second, as in Figure \ref{fig:Tsurf1}, Figure \ref{fig:Tsurf2} shows
that the RC model overpredicts nightside surface temperatures (top
left panel). In contrast, the RCS model fits the GCM values extremely
well (bottom left panel). For nightside temperatures we find
$r^2=0.76$ with the RC model, while the RCS model essentially
reproduces the GCM results with a fit of $r^2 = 0.98$.

To conclude, we have formulated a radiative-convective-subsiding model that
utilizes WTG balance combined with the heat engine scaling for the
large-scale circulation to capture the day-night heat
transport and nightside temperature structure. Our model captures the
day-night temperature structure of many GCM simulations extremely well.
We provide an intuitive understanding of the model results in the
next section.

\section{Transition to large day-night temperature gradients}
\label{sec:transition}

In this section we explain the threshold at which atmospheres of
tidally locked rocky planets develop large day-night temperature
gradients. We point out again that hot Jupiter theories suggest this
should occur when $t_{wave}/t_{rad} \gtrsim 1$ (Section
\ref{sec:introduction}). In contrast, we show that on rocky planets
the threshold is up to two orders of magnitude smaller and temperature
gradients become large when
$t_{wave}/t_{rad} \gtrsim \mathcal{O}(10^{-2})$.
The small threshold
is important because it means rocky exoplanets are relatively more
sensitive to the parameter $t_{wave}/t_{rad}$, so planets that are
relatively cool or have thick atmospheres still exhibit large
day-night temperature differences.
Finally, we relate the RCS model back to previous theories by showing
that it reduces to our RC model for $t_{wave}/t_{rad}\lesssim10^{-4}$ and to
\citet{wordsworth2015}'s result for $t_{wave}/t_{rad}\gtrsim1$ and $\tau_{LW}\ll1$.

To start, we consider the thermodynamic equation under WTG balance (Equation \ref{eqn:rcs1}a). WTG balance expresses a balance between subsidence heating and
radiative cooling, and we nondimensionalize it
using $\hat{T} = T/T_{eq}$ and $\hat{F} = F/(\sigma T_{eq}^4)$. We
find that the ratio of subsidence heating to radiative cooling is
governed by two parameters,
\begin{eqnarray}
 \frac{d \hat{T}}{d \tau} - \beta \frac{\hat{T}}{\tau}&=& \left( \frac{t_{sub}}{t_{rad}}  \right)\frac{d \hat{F} }{d \tau},
\label{eqn:wtg}
\end{eqnarray}
where $\beta=R/(c_p n)$ sets the adiabatic lapse rate, $t_{sub} \equiv
p_s/\bar{\omega}$ is a characteristic subsidence timescale for a parcel of air and
$t_{rad} = p_s c_p /(g \sigma T_{eq}^3)$ is the radiative cooling
timescale. Equation \ref{eqn:wtg} is the three-dimensional equivalent of the
WTG scaling developed by \citet{perez-becker2013a} using the shallow-water
equations.
The lapse rate parameter $\beta$ is always of order unity whereas the subsidence timescale $t_{sub}$ is an emergent
timescale set by the large-scale dynamics.
When $t_{sub}/t_{rad} \ll 1$ radiative cooling is inefficient compared
with subsidence heating, the nightside atmosphere is close to an
adiabat, and day-night temperature differences are small. When $t_{sub}/t_{rad} \gtrsim 1$ a parcel of air cools
significantly as it descends, the nightside develops inversions, and
day-night differences are large. Finally, for
$t_{sub}/t_{rad} \gg 1$ the nightside is close to radiative
equilibrium.

%%%%
\begin{figure}[t!]
%\figurenum{10}
%\plotone{f10.eps}
\plotone{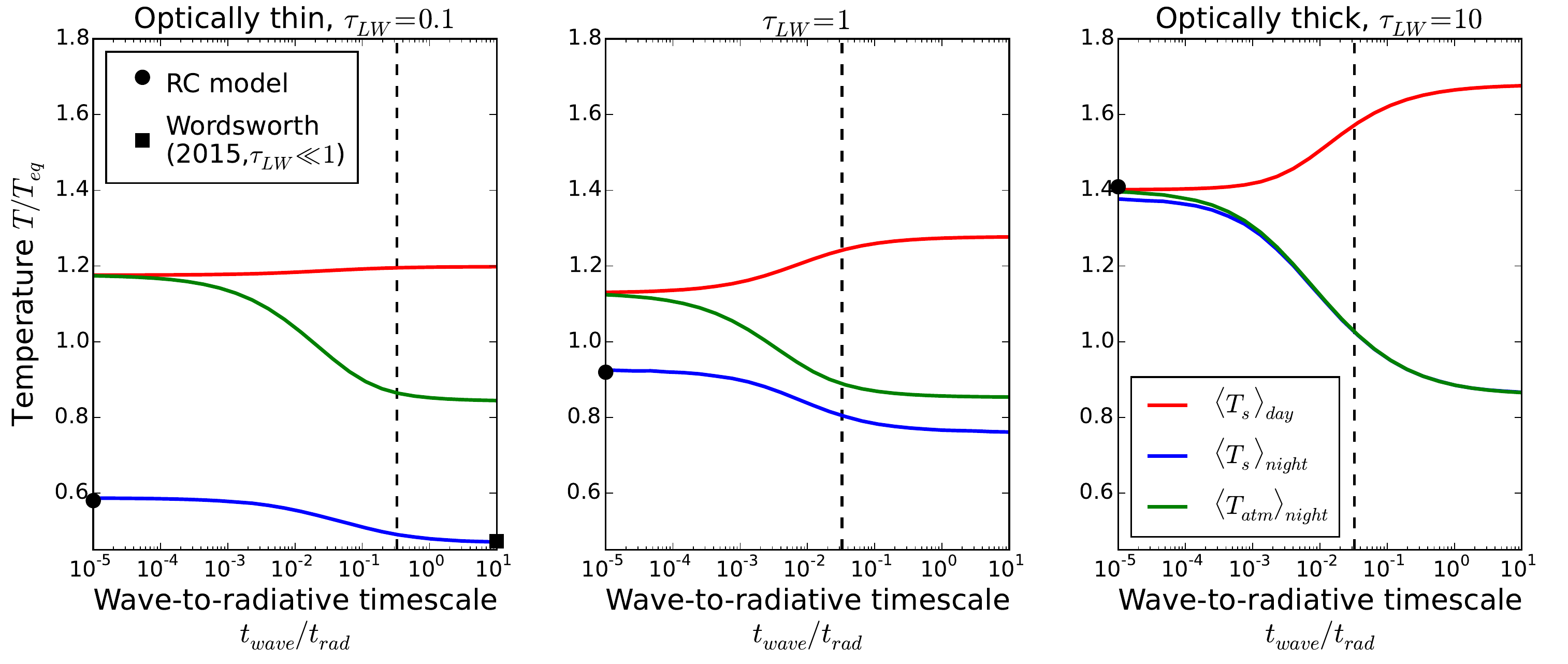}
  \caption{Day-night temperature gradients are large once the
    wave-to-radiative timescale ratio $t_{wave}/t_{rad}$ exceeds the
    threshold from Equation \ref{eqn:transition3} (vertical dashed
    lines). The panels show dayside
    surface temperature, $\langle T_s\rangle_{day}$, nightside surface
    temperature, $\langle T_s\rangle_{night}$, and the bottom-most
    atmospheric temperature on the nightside,
    $\langle T_{atm}\rangle_{night}$, from the
    radiative-convective-subsiding model (RCS, Section \ref{sec:radsub}).
    In all cases, cool/thick atmospheres with
    $t_{wave}/t_{rad} \lesssim 10^{-4}$ have small temperature
    gradients between dayside surface and nightside
    atmosphere. Surface temperature gradients additionally
    depend on optical thickness, and even cool/thick atmospheres can
    have large day-night surface temperature gradients if
    $\tau_{LW}\ll 1$ (left panel). Black symbols show the nightside surface
    temperatures predicted by the radiative-convective model (RC), and
    the asymptotic scaling of \citet{wordsworth2015}; the RCS
    model reduces to either in the limits $t_{wave}/t_{rad} \lesssim
    10^{-4}$ and $t_{wave}/t_{rad} \gtrsim 1$.
  \label{fig:dTemp1}}
\end{figure}
%%%%

The transition to large day-night temperature gradients occurs at
a wave-to-radiative timescale threshold of
$t_{wave}/t_{rad} \sim 10^{-2}$.  Figure \ref{fig:dTemp1} shows
temperatures from the RCS model as a function of the timescale ratio
$t_{wave}/t_{rad}$ and optical thickness $\tau_{LW}$.  We assume a
representative rocky planet scenario\footnote{We assume an Earth-sized
  planet, $a=a_\Earth$, with an N$_2$-dominated atmosphere,
  $(R,c_p) = (R,c_p)_{N_2}$.}, and plot the dayside surface
temperature, $\langle T_s\rangle_{day}$, nightside surface
temperature, $\langle T_s\rangle_{night}$, and the atmospheric
temperature just above the nightside surface,
$\langle T_{atm} \rangle_{night}$. Because the atmospheric temperature
on the dayside is strongly coupled to the surface via convection,
$\langle T_{atm}\rangle_{day}\approx \langle T_s\rangle_{day}$, the
difference between $\langle T_s\rangle_{day}$ and
$\langle T_{atm}\rangle_{night}$ also shows the day-night temperature
gradient in the lowest part of the atmosphere.  First, it is clear
from Figure \ref{fig:dTemp1} that the atmospheric temperature gradient
is small when $t_{wave}/t_{rad} \ll 10^{-2}$. The transition to large
temperature gradients spans many orders of magnitude, but we take
$t_{wave}/t_{rad} \sim 10^{-2}$ as a representative value that ensures
temperature gradients are large for larger values of
$t_{wave}/t_{rad}$. 
Second, once day-night atmospheric gradients are large their magnitude 
  additionally depends on $\tau_{LW}$, with optically thicker
  atmospheres having larger maximum temperature gradients
  (compare maximum difference between $\langle T_s\rangle_{day}$ and $\langle T_{atm} \rangle_{night}$).
Third, because the nightside surface is in
radiative equilibrium with the overlying atmosphere, the gradient in
surface temperatures is at least as big as the gradient in
atmospheric temperatures. However, it can be much larger in the
optically thin limit because the nightside atmosphere becomes
ineffective at radiatively heating the nightside surface. At low
optical thickness the nightside surface is much colder than the
overlying air (Fig.~\ref{fig:dTemp1}a), while at high optical
thickness the nightside surface is closely tied to the overlying air
temperature (Fig.~\ref{fig:dTemp1}c).

Next, we explain why atmospheres develop large temperature gradients
at $t_{wave}/t_{rad}\sim 10^{-2}$.  As we showed above, the nightside
temperature structure is controlled by the ratio of
subsidence to radiative timescales, $t_{sub}/t_{rad}$. Here we
analyze the processes that control
$t_{sub}$. Using the heat engine and the area ratio between upward and
downward motions, we already found the pressure velocity on the
nightside $\omega_{down}$. Equation \ref{eqn:rcs6} allows us to write
\begin{eqnarray}
  t_{sub} = \frac{p_s}{\omega_{down}} & = & \frac{a}{\chi U_s}.
\label{eqn:transition1}
\end{eqnarray}
Next, we scale the surface wind speed $U_s$ from the heat engine (Equation \ref{eqn:he1})
\begin{eqnarray}
 U_s & = & \left[ (T_{d}/T_{eq} - 1) \times (1-e^{-\tau_{LW}})
                       \left(\frac{c_p}{R}\right)^2
                       \frac{t_{drag}}{t_{rad}} \right]^{1/3} \times
                       c_{wave}, \nonumber \\
  U_s & \approx & \left( \frac{c_p}{R} \right)^{2/3}
                       \left( \frac{t_{drag}}{t_{rad}} \right)^{1/3}
                          \times c_{wave},
\label{eqn:transition2}
\end{eqnarray}
where in the second step we assumed an optically thick atmosphere,
$\tau_{LW}\geq1$, so that all incoming stellar flux goes towards
driving atmospheric motion, $1-e^{-\tau_{LW}}\approx 1$. We also drop
the dependence on the dayside temperature from $T_d/T_{eq}-1$. We do
so because in the optically thick limit $T_d$ is approximately given
by Equation \ref{eqn:pierrehumbert1}, so
$(T_d/T_{eq}-1)^{1/3} \approx (\tau_{LW}^{\beta}
\Gamma[1+4\beta]^{-1/4} - 1)^{1/3}$
which is always of order unity\footnote{For example, assuming
  $\tau_{LW}=2$ and a diatomic gas without pressure broadening
  ($\beta=2/7$), $(T_d/T_{eq}-1)^{1/3} \approx 0.6$. The gamma
  function $\Gamma[1+4\beta]^{-1/4}$ does not vary significantly over
  the plausible range of atmospheric gases. Similarly, the dependency
  on $\tau_{LW}^{\beta/3}$ is negligible because the exponent
  $\beta/3$ is always small.}.  We combine Equations
\ref{eqn:transition1} and \ref{eqn:transition2} and find
\begin{eqnarray}
  t_{sub} & = & \frac{1}{\chi} \left( \frac{R}{c_p} \right)^{2/3}
                \left( \frac{t_{rad}}{t_{drag}} \right)^{1/3} t_{wave}
\label{eqn:x2}
\end{eqnarray}
Equation \ref{eqn:x2} gives us the subsidence time on the nightside.
Day-night temperature gradients will be small if a parcel of
air cools slower than it sinks, $t_{rad} >
t_{sub}$. Conversely, day-night temperature gradients will be large if
a parcel cools faster
than it sinks, $t_{rad} < t_{sub}$. The threshold between these two
regimes is
\begin{eqnarray}
  t_{rad} & \sim & t_{sub} \nonumber \\
  t_{rad} & \sim & \frac{1}{\chi} \left( \frac{R}{c_p} \right)^{2/3}
                   \left( \frac{t_{rad}}{t_{drag}} \right)^{1/3} t_{wave} \nonumber\\
  t_{rad}^{2/3} & \sim & \frac{1}{\chi} \left( \frac{R}{c_p} \right)^{2/3}
                   \left( \frac{1}{t_{drag}} \right)^{1/3} t_{wave} \nonumber\\
  t_{rad} & \sim & \left(\frac{1}{\chi}\right)^{3/2} \left( \frac{R}{c_p} \right)
                   \left( \frac{t_{wave}}{t_{drag}} \right)^{1/2}
                   t_{wave} ~(\text{for } \tau_{LW} \geq 1).
\end{eqnarray}

We can find a similar threshold for optically thin atmospheres
($\tau_{LW}<1$). We note that the standard radiative timescale
$t_{rad} = c_p p_s /(g \sigma T_{eq}^3)$ is the cooling timescale of
an \textit{optically thick} column of air. In contrast, an optically
thin column of air only emits a radiative flux
$\sim \tau_{LW}\times \sigma T_{eq}^4$ so its radiative cooling
timescale is
\begin{eqnarray}
  t_{rad,thin} & = & \frac{t_{rad}}{\tau_{LW}}.
\label{eqn:tradthin}
\end{eqnarray}
WTG balance (Equation \ref{eqn:wtg}) in the optically thin regime is
still governed by the ratio of subsidence to radiative timescales, but
now $t_{rad}$ has to be replaced by $t_{rad,thin}$.

To find the subsidence timescale $t_{sub}$ in the optically thin limit, we note that optically
thin atmospheres are also less efficient heat engines.
The lower efficiency arises because, for $\tau_{LW}\ll1$, the surface
re-emits most of the incoming stellar flux directly back to space and only a
fraction $1-e^{-\tau_{LW}} = 1-(1-\tau_{LW}+...)\approx \tau_{LW}$ of
the stellar flux is available to drive atmospheric motions. The
dayside temperature $T_d$ is approximately constant in the optically
thin case (Fig.~\ref{fig:Tsurf1}), so $t_{sub}$ is
\begin{eqnarray}
  t_{sub} & = & \frac{1}{\chi} \left( \frac{R}{c_p} \right)^{2/3}
                \left( \frac{t_{rad,thin}}{t_{drag}} \right)^{1/3} t_{wave}.
\label{eqn:x3}
\end{eqnarray}
Equation \ref{eqn:x3} only differs from Equation \ref{eqn:x2} through
the use of $t_{rad,thin}$ instead of $t_{rad}$. Our result for large
temperature gradients therefore also holds for optically thin
atmospheres, once we replace $t_{rad}$ with $t_{rad,thin}$.

To compare our result with the result for hot Jupiters, we express
the criterion for an atmosphere to develop large temperature gradients in terms of the wave-to-radiative timescale ratio
$t_{wave}/t_{rad}$. Day-night atmospheric temperature gradients are large once
\begin{eqnarray}
  \frac{t_{wave}}{t_{rad}} & \gtrsim & 
                                       \begin{cases} \chi^{3/2} \times \dfrac{c_p}{R}
                                       \left( \dfrac{t_{drag}}{t_{wave}} \right)^{1/2}&
                                       ~\text{if } \tau_{LW} \geq 1,\\ 
                                       \\
                                       \dfrac{\chi^{3/2}}{\tau_{LW}} \times \dfrac{c_p}{R}
                                       \left(
                                         \dfrac{t_{drag}}{t_{wave}} \right)^{1/2} &
                                       ~\text{if } \tau_{LW} < 1.
                                       \end{cases}
\label{eqn:transition3}
\end{eqnarray}
We emphasize that Equation \ref{eqn:transition3} only ensures that
atmospheric temperature gradients are large, but they remain significant until
$t_{wave}/t_{rad}$ becomes extremely small (Fig.~\ref{fig:dTemp1}).

We draw three important conclusions from Equation
\ref{eqn:transition3}. First, in the optically thick case the right
hand side is dominated by $\chi \approx 1/20$ (Section \ref{sec:radsub}) while the other
quantities do not vary much in most cases of interest. The small value
of $\chi$ causes the
threshold for large day-night temperature gradients to generally be
much smaller than one.
As a representative high mean-molecular-weight (MMW) scenario, we
consider an N$_2$ atmosphere with $T_{eq}=300$ K. In this case
$c_p/R = 7/2$ and $t_{drag}/t_{wave} = 1.4$ (Appendix
\ref{sec:appendix1}), so temperature gradients are large when
\begin{eqnarray}
  \left( \frac{t_{wave}}{t_{rad}} \right)_{high~MMW}& \gtrsim & 5 \times 10^{-2}.
\end{eqnarray}
Our result explains why rocky planets develop large atmospheric temperature
gradients at a threshold almost two orders of magnitudes smaller than
what one would expect based on the results for hot Jupiters, $t_{wave}/t_{rad}\gtrsim1$.

Second, hot H$_2$-dominated atmospheres are a notable exception to
the first result and develop day-night temperature gradients at
larger values of $t_{wave}/t_{rad}$. The larger threshold arises because
H$_2$-dominated atmospheres have larger scale heights than
high-MMW atmospheres, which increases the drag time $t_{drag}$. For
example, we consider a H$_2$ atmosphere with $T_{eq} = 600$ K. In this
case $t_{drag}/t_{wave} = 40$ (Appendix
\ref{sec:appendix1}) so temperature gradients are large when
\begin{eqnarray}
  \left( \frac{t_{wave}}{t_{rad}} \right)_{H_2}& \gtrsim & 0.2.
\end{eqnarray}
The wave-to-radiative timescale threshold in this case is a factor of
four larger than for high-MMW atmospheres, but it is still
almost an order of magnitude smaller than the result for hot Jupiters,
$t_{wave}/t_{rad}\gtrsim1$.

Third, optically thin atmospheres are less prone to developing
day-night temperature gradients than optically thick atmospheres
because optically thin atmospheres cool less effectively. Although
optically thin atmospheres are also less efficient heat engines, the
radiative effect dominates because $t_{rad,thin} \propto 1/\tau_{LW}$
whereas
$t_{sub} \propto t_{rad,thin}^{1/3} \propto 1/\tau_{LW}^{1/3}$.
Weak-temperature-gradient (WTG) balance therefore holds even better in
optically thin atmospheres than in optically thick ones.  It also
explains why the stratospheres of our simulations, where the
atmosphere becomes optically thin \citep{robinson2012}, are much more
horizontally homogeneous than the lower atmosphere
(Fig.~\ref{fig:climDry50a}).

We can now relate the RCS model to the results in Section
\ref{sec:radconv}. First, when $t_{sub}/t_{rad}\ll1$ radiative cooling
is inefficient compared to subsidence heating. In this limit sinking
parcels of air on the nightside remain close to an adiabat and the RCS
model reduces to the RC model\footnote{Because the RCS model
  additionally includes a stratosphere, it predicts slightly colder
  nightsides in the limit $t_{wave}/t_{rad}<10^{-4}$ than the RC model
  but the effect is small (the black dot is slightly above the
    blue line in Figure \ref{fig:dTemp1}, right
    panel).}. Figure
\ref{fig:dTemp1} shows nightside surface temperatures in both models
and demonstrates that the RCS model reduces to the RC model at a
representative value of $t_{wave}/t_{rad} \lesssim 10^{-4}$ (compare
blue lines to black dots).
Second, when $t_{sub}/t_{rad}\gg1$ radiative cooling is much stronger
than subsidence heating. In this limit WTG balance
(Eqn.~\ref{eqn:wtg}) becomes $d F/d\tau \approx 0$, so the nightside
is in purely radiative equilibrium and $F$ is vertically constant. To
still satisfy the nightside surface budget (Equation \ref{eqn:rcs4}c),
$F$ has to be zero. The Schwarzschild equation (Eqn.~\ref{eqn:rcs1}b)
shows that in this case $d(\sigma T^{4})/d\tau \approx 0$ which means
the nightside becomes vertically isothermal with a temperature that is
set by the overlying tropopause temperature $T_{strat}(\tau_0)$. A
lower bound for $T_{strat}$ (see Eqn.~\ref{eqn:rcs2}) is given by the
skin temperature $T_{skin}\equiv2^{-1/4} T_{eq}$
\citep{pierrehumbert2011b}. In the optically thin limit
the nightside surface energy budget is then equal to
$\sigma T_n^4 = \tau_{LW} \times \sigma T_{skin}^4 =
\tau_{LW} \times \sigma T_{eq}^4/2$,
and we recover \citet{wordsworth2015}'s result
$T_n = T_{eq} (\tau_{LW}/2)^{1/4}$.  Figure \ref{fig:dTemp1} shows
that the nightside temperature in the RCS model reduces to this limit
at a representative value of $t_{wave}/t_{rad} \gtrsim 1$ (compare black
square and blue line in left panel).

Up to now we have focused on slowly rotating planets $a^2/L_{Ro}^2 <
1$. Next, we consider the effects of rapid rotation, and how they
interact with the threshold for large day-night temperature gradients.

\clearpage

\section{Effects of rapid rotation on temperature structure}
\label{sec:rotation}

In this section we use GCM simulations to address how rapid rotation
affects the circulation and temperature structure. \citet{leconte2013}
showed that tidally locked planets develop drastically different
circulations when $a^2/L_{Ro}^2 \gtrsim 1$ because equatorial waves
are not able to freely propagate into high latitudes once the
planetary radius, $a$, is larger than the equatorial Rossby radius,
$L_{Ro}$. Rapidly rotating planets then develop standing Rossby and
Kelvin wave patterns. The standing wave patterns lead to strong
equatorial superrotation, an eastward offset of the equatorial hot
spot, and off-equatorial cold vortices on the nightside
\citep{matsuno1966,showman2011}.  Here we also find that the
circulation regime changes at $a^2/L_{Ro}^2 \sim 1$. However, while
rapid rotation drastically alters the flow field the
effect on temperature structure is small unless the atmosphere is also
prone to developing strong temperature gradients,
$t_{wave}/t_{rad} \gtrsim \mathcal{O}(10^{-2})$.

We perform a set of GCM simulations in which we vary $a^2/L_{Ro}^2$
and $t_{wave}/t_{rad}$ while keeping all other parameters fixed. We
vary $a^2/L_{Ro}^2$ by changing the rotation rate $\Omega$, and
$t_{wave}/t_{rad}$ by changing the surface pressure $p_s$. All other
parameters are fixed to the same values as the reference simulation in Figure
\ref{fig:climDry50a}, that is, $a=a_\Earth$, $T_{eq}=283$K,
$(R,c_p)=(R,c_p)_{N_2}$, and $\tau_{LW}=1$. We explore
$a^2/L_{Ro}^2=(0.1,0.5,1)$ and
$t_{wave}/t_{rad}=(10^{-3},10^{-2},10^{-1})$, which correspond to
$2\pi/60$ days $\leq \Omega \leq 2\pi/6$ days and $5$ bar $\leq p_s
\leq 0.05$ bar.

%%%%
\begin{figure}[t!]
%\figurenum{11}
%\plotone{f11.eps}
\plotone{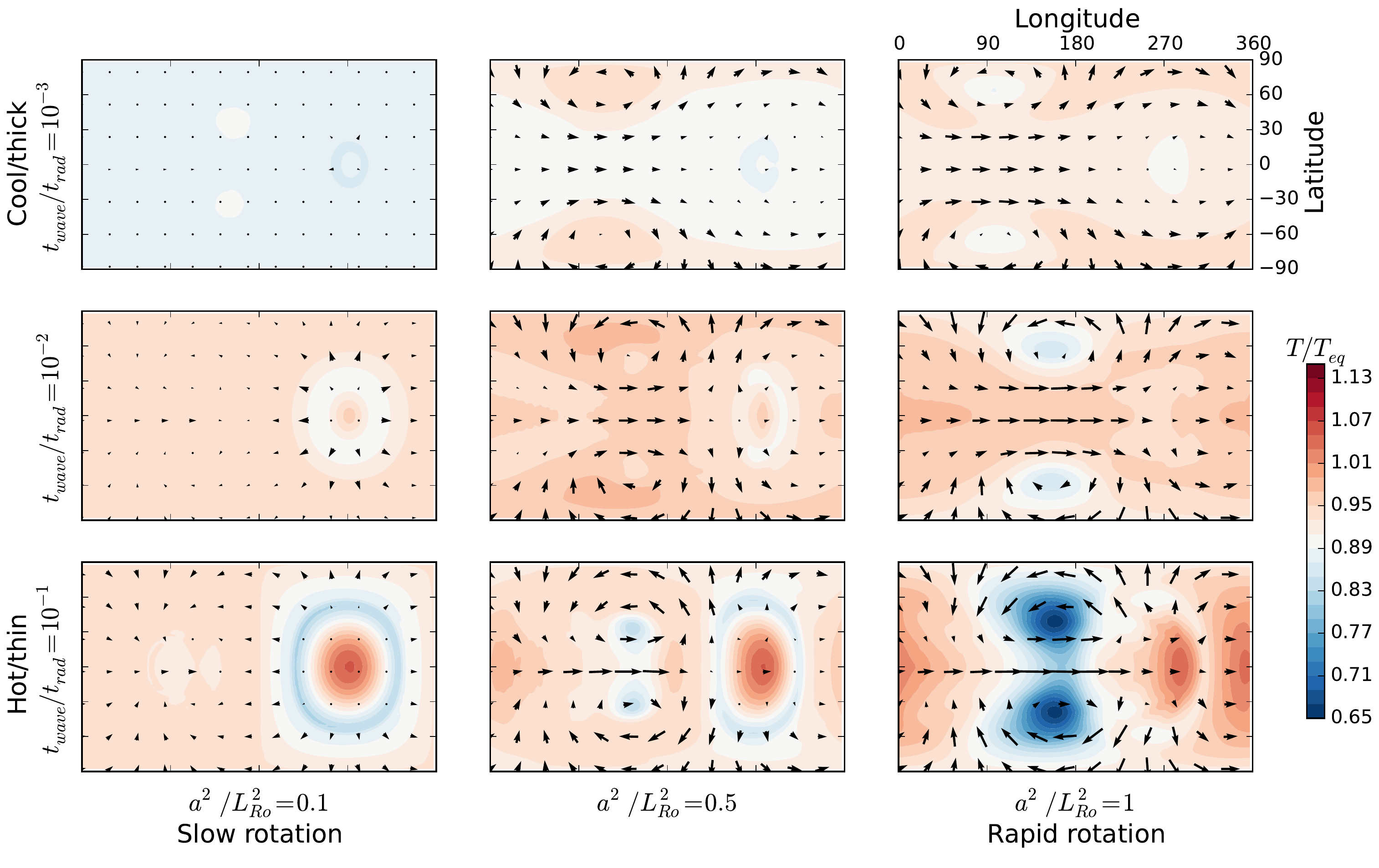}
  \caption{Rapid rotation ($a^2/L_{Ro}^2\gtrsim1$) does not have a strong
    effect on temperature structure unless the atmosphere is also
    hot or thin ($t_{wave}/t_{rad}>10^{-2}$). Shown are 2D temperature and
    wind fields, averaged over the upper troposphere
    ($0.3\leq p/p_s\leq0.4$).  Rotation increases from left to right,
    the wave-to-radiative timescale ratio increases from top to
    bottom. Increased rotation changes the circulation drastically,
    from a day-night flow at slow rotation (left) to an equatorially
    superrotating jet and cold nightside vortices at rapid rotation
    (right). However, at low $t_{wave}/t_{rad}$ temperature gradients
    are small, even if rotation is rapid (top right). Large temperature
    gradients, eastward hot spot offsets, and cold nightside vortices
    only emerge once an atmosphere is both hot/thin and rotates rapidly (bottom
    right). The substellar point is located at $270^\circ$ longitude.\label{fig:Ttrop1}}
\end{figure}
%%%%

We find that rapid rotation has a large effect on the
circulation, but its effect on the temperature structure is small
unless $t_{wave}/t_{rad}$ also exceeds the threshold $t_{wave}/t_{rad}
\gtrsim 5\times 10^{-2}$ from the previous section.
Figure \ref{fig:Ttrop1} shows 2D maps of the circulation
and temperatures in the upper atmosphere. The wind and
temperatures are mass-weighted averages taken over
$0.3 \leq p/p_s \leq 0.4$, and the substellar point is located at
longitude $\lambda = 270^{\circ}$.
Slowly rotating simulations are shown in the left column of Figure
\ref{fig:Ttrop1}. As expected, the circulation consists of a
substellar-to-antistellar flow. At small values of $t_{wave}/t_{rad}$ the day-night
temperature differences are small, but the atmosphere develops large
day-night temperature gradients at $t_{wave}/t_{rad}=10^{-1}$,
consistent with our results in Section \ref{sec:transition}.
The top row of Figure \ref{fig:Ttrop1} shows simulations with small $t_{wave}/t_{rad}$. As rotation rate increases, the atmospheric circulation
changes drastically. A strong equatorial jet develops and the
nightside atmosphere additionally develops standing Rossby waves in
the form of off-equatorial vortices
\citep{showman2011}. Nevertheless, as long as
$t_{wave}/t_{rad}=10^{-3}$, the maximum horizontal temperature
difference at
$a^2/L_{Ro}^2 = 1$ only reaches $0.05 T_{eq}$, or $\sim15$ K.

Figure \ref{fig:Ttrop1} underlines that day-night atmospheric
temperature gradients are primarily controlled by
$t_{wave}/t_{rad}$. However, the effect of rapid rotation can strongly
enhance temperature gradients in the form of eastward hot spot
offsets and cold nightside vortices. The slowly rotating simulation
with a thin atmosphere ($a^2/L_{Ro}^2 = 0.1$,
$t_{wave}/t_{rad}=10^{-1}$; bottom left) has its hottest point located
at the substellar point and a maximum horizontal temperature
difference of $0.2 T_{eq}$, or $\sim65$ K.  In contrast, the
simulation with the same value of $t_{wave}/t_{rad}$ but at rapid
rotation shows an eastward hot spot offset and a significantly
larger maximal horizontal temperature difference of $0.4 T_{eq}$, or
$\sim110$ K (bottom right).
In the following section we consider what our results imply for future
observations.

\section{Implications for Observations}
\label{sec:obs}

%%%%
\begin{figure}[b!]
%\figurenum{12}
%\plotone{f12.eps}
\plotone{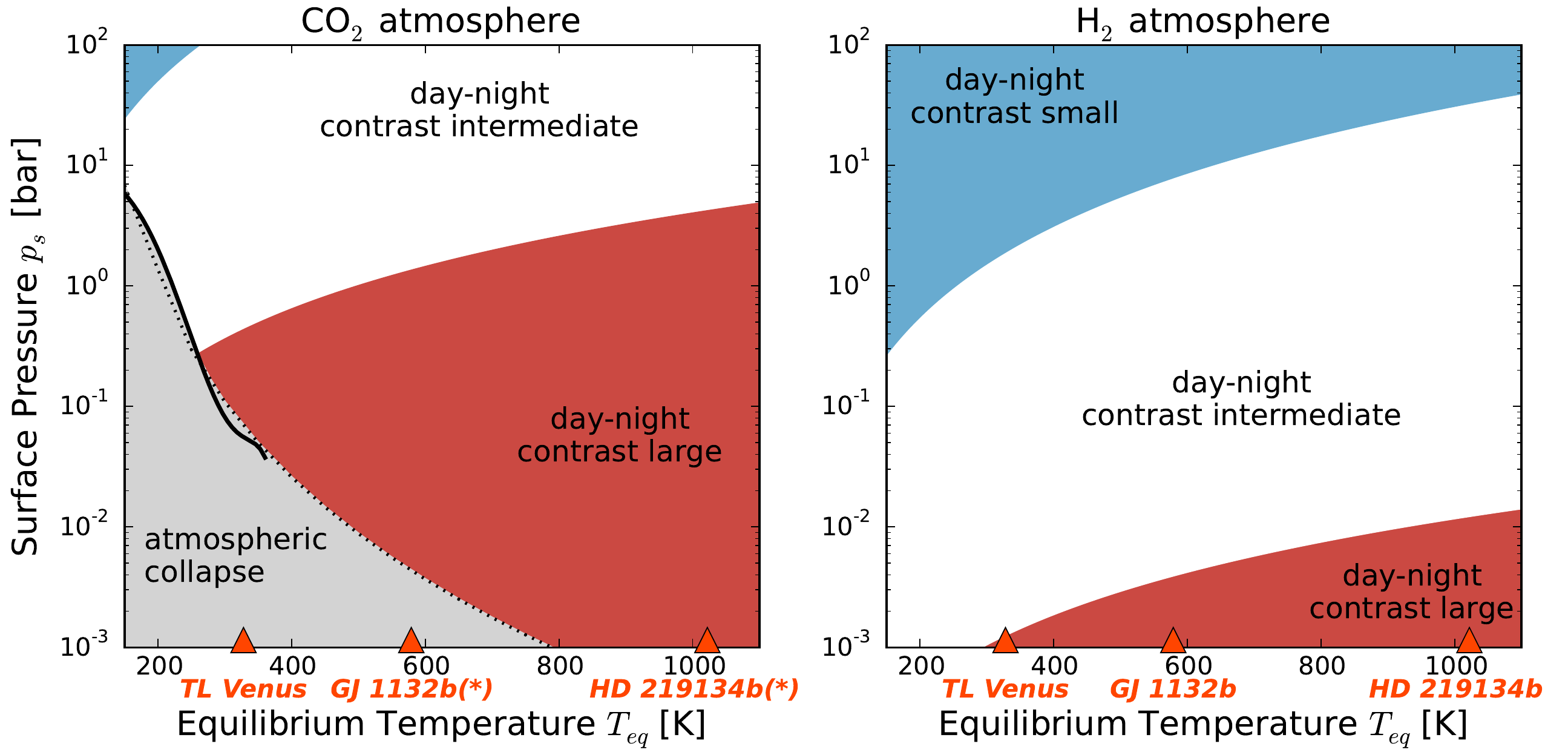}
  \caption{CO$_2$ atmospheres are more likely to develop large
    temperature gradients than H$_2$ atmospheres.  Atmospheric
    day-night temperature gradients are negligible inside the blue
    region ($t_{wave}/t_{rad}\leq10^{-4}$) and are large inside the
    red region (Eqn.~\ref{eqn:transition3} for $\tau_{LW}\geq1$).
    CO$_2$ atmospheres collapse inside the grey region [solid line:
    empirical fit to GCM results from \citet{wordsworth2015}; dotted
    line: calculated using our RCS
    model]. Bottom symbols show equilibrium temperatures of
    two nearby rocky planets and of a hypothetical tidally locked
    Venus; $(*)$ marks scenarios for which rotational effects would
    additionally be important ($a^2/L_{Ro}^2\geq1$). The shown
    thresholds assume a GJ1132b-sized planet ($a,g=1.16a_\Earth,11.7$
    m s$^{-2}$).\label{fig:obs}}
\end{figure}
%%%%
Figure \ref{fig:obs} summarizes some implications of our results for
observations of rocky exoplanets.  We showed that atmospheric
day-night temperature contrasts strongly depend on the parameter
$t_{wave}/t_{rad} = a/(\sqrt{R/c_p} \times \sqrt{R T_{eq}}) \times g
\sigma T_{eq}^3 /(c_p p_s) \propto T_{eq}^{5/2} / p_s$.
The planetary radius $a$ and surface gravity $g$ vary relatively
little for plausible rocky planets, which means day-night temperature
differences are to first order controlled by the equilibrium
temperature $T_{eq}$, the surface pressure $p_s$, and whether or not
the atmosphere is made of H$_2$ (via $R$ and $c_p$).
In Figure \ref{fig:obs} we consider these parameters for a GJ 1132b-sized planet\footnote{We assume
  $a,g=1.16a_\Earth,11.7$ m s$^{-2}$ \citep{berta-thompson2015}.} with
hypothetical CO$_2$ and H$_2$ atmospheres.
The red region indicates when the atmosphere is hot/thin and develops
large day-night atmospheric temperature gradients (Equation
\ref{eqn:transition3} for $\tau_{LW}\geq1$). The blue region indicates
when the atmosphere is cool/thick and day-night atmospheric
temperature gradients become negligible ($t_{wave}/t_{rad}\leq10^{-4}$
from Section \ref{sec:transition}).
Part of the CO$_2$ phase space is unstable to atmospheric collapse,
which occurs when the nightside surface is cold enough for CO$_2$ to
condense.  We delineate atmospheric collapse using two
approaches. First, the solid black line shows the empirical fit from
\citet{wordsworth2015}, who used a GCM with full radiative transfer to
compute collapse thresholds up to $T_{eq}=367$ K. Second, we use the
RCS model to compute when nightside surface temperatures fall below
the condensation temperature of CO$_2$. To specify the optical
thickness we use Equation \ref{eqn:powerlaw} and assume $\tau(1$
bar$)=1$ and $n=1$.  Although we do not include non-grey effects, the
RCS model (dashed black line) closely fits the GCM results (solid
black line) over the range of parameters explored by
\citet{wordsworth2015}. As such, we consider the RCS model appropriate
for predicting atmospheric collapse (also see Section
\ref{sec:discussion}). We repeat a similar computation for
H$_2$ atmospheres, and find that the entire phase space in Figure
\ref{fig:obs}b is stable against collapse\footnote{We use the Solar
  opacity value in \citet{menou2012a} and $n=1$, and compute nightside
  surface temperatures with the RCS model. We find that nightside
  temperatures always exceed the critical point of H$_2$, 33.2 K.}.
The symbols at the bottom of Figure \ref{fig:obs}
show equilibrium temperatures\footnote{We assume a planetary albedo of
  zero.} of two
recently-discovered rocky planets and of a hypothetical tidally locked
planet at Venus' present-day orbit
\citep{berta-thompson2015,motalebi2015}.
Finally, we found that temperature structure can be affected by
rapid rotation. We mark all rapidly rotating
planet scenarios with $a^2/L_{Ro} \geq 1$ using star symbols ($*$). For these
cases we expect that strong rotational effects, such as large eastward
hot spot offsets or cold nightside vortices, occur
inside the red region (see Section \ref{sec:rotation}).

Figure \ref{fig:obs} allows us to make some tentative
predictions. First, with
a high MMW atmosphere like CO$_2$, GJ 1132b and HD 219134b would have
non-negligible day-night temperature gradients. This conclusion holds
even for surface pressures as high as that of Venus ($p_s=$ 92
bar). Second, GJ 1132b and HD 219134b with CO$_2$ atmospheres both
satisfy the criterion for rapid rotation ($a^2/L_{Ro}^2 \geq 1$).
Should observations detect a large eastward hot spot offset, it would
favor surface pressures less than $\mathcal{O}(1)$ bar (inside the red
region).  Third, Figure \ref{fig:obs}a shows that a CO$_2$ atmosphere
with surface pressure comparable to that of Mars
($p_s=6\times 10^{-3}$ bar) would be close to collapse on GJ 1132b. We
note that our collapse calculation does not account for rotational
effects, and cold nightside vortices (Fig.\ref{fig:Ttrop1}) would
allow the atmosphere to collapse at even higher pressures. Fourth, if
these planets managed to retain H$_2$-dominated atmospheres against
atmospheric escape, they would be stable against collapse and exhibit
much smaller day-night temperature differences than similar CO$_2$
atmospheres. The increased stability and smaller temperature gradient
is due to a combination of H$_2$'s large heat capacity $c_p$, which
increases the radiative timescale $t_{rad}$ \citep[][]{menou2012a},
its large gas constant $R$, which increases the speed of atmospheric
waves $c_{wave}$ and thus decreases the wave timescale $t_{wave}$
\citep[][]{heng2012a}, and its increased scale height, which decreases
the effect of friction (Appendix \ref{sec:appendix1}).  Fifth,
H$_2$-dominated atmospheres would be significantly less affected by
rotation than CO$_2$ atmospheres. The smaller effect of rotation is
also due to the reduced wave timescale $t_{wave}$ in H$_2$
atmospheres. For example, assuming GJ 1132b's equilibrium temperature
$T_{eq}=579$K, the characteristic speed of gravity waves in a CO$_2$
atmosphere is $c_{wave} = \sqrt{R/c_{p}} \times\sqrt{R T_{eq}} =158$m
s$^{-1}$, whereas in a H$_2$ atmosphere $c_{wave} = 838$m s$^{-1}$. It
follows that the nondimensional Rossby radius,
$a^2/L_{Ro}^2 = 2 \Omega a/c_{wave}$, is about five times smaller in a
H$_2$ atmosphere. Our results also imply that rocky planets with
H$_2$-dominated atmospheres are less likely to exhibit eastward hot
spot offsets and cold nightside vortices.
These predictions are qualitative because they do not consider
  the optical thickness $\tau_{LW}$, which helps set the magnitude
  of the day-night temperature gradient
  (Fig.~\ref{fig:dTemp1}). Quantitatively interpreting an observed
  day-night temperature gradient also requires constraining
  $\tau_{LW}$, for example via transit spectroscopy \citep[see][]{koll2015}.

One way of distinguishing the scenarios in Figure \ref{fig:obs} would
thus be through combined transit spectroscopy and
thermal phase curve observations with \textit{JWST}. Previous
feasibility studies have tended to emphasize the transit technique
\citep[e.g.,][]{beichman2014,batalha2015a}, here we compare the
signal-to-noise ratio (SNR) that can be achieved by spending the same
amount of \textit{JWST} time on low spectral-resolution transit and
broadband phase curve observations.  We find that it would
  take about as much time to measure the broadband mid-IR
  phase curve of a short-period rocky exoplanet as it would to detect
  molecular signatures in its atmospheres through near-IR
  transit spectroscopy.
The basic science goal for transit observations would be to detect a
molecular species from its spectral imprint; a flat spectrum could
alternately be a cloudy atmosphere or no atmosphere. The basic science
goal for phase curve observations would be to detect the day-night
flux difference of a bare rock. An observed flux difference lower than
that of a bare rock would imply the presence of an atmosphere thick
enough to modify the day-night temperature contrast (outside the red
region in Fig.~\ref{fig:obs}). Similarly, hot/cold spot offsets would
imply the presence of an atmosphere that is hot or thin enough that
its thermal structure is significantly affected by rotation (see
above).
We consider GJ 1132b with a CO$_2$-dominated atmosphere
as a representative target. We assume GJ 1132b is tidally locked,
which could be verified using optical phase curves with
\textit{TESS} \citep[cf.][]{fujii2014}.
We compute transit signals following \citet{cowan2015}, but assume that
spectral features in the near-IR cause an absorption difference of three
scale heights and have a typical width of $0.1 \mu$m
\citep[see Table 2,][]{kaltenegger2009}. We compute the phase curve signal of a bare rock
following \citet{koll2015}.  We estimate \textit{JWST}'s precision in
the near-IR ($1-4\mu$m in $0.1 \mu$m bins, $R\sim25$ on NIRSpec) using
the photon noise limit \citep[see][]{koll2015}.  We similarly estimate
the precision in the mid-IR ($16.5-19.5\mu$m broadband, F1800W
on MIRI) assuming photon noise, but account for the imperfect
instrument throughput of 1/3 \citep{glasse2010}. We assume that both
techniques bin photons over the length of one transit
($45$ minutes) and we multiply the noise by $\sqrt{2}$ to account for the
fact that both techniques compare two
snapshots in time. We assume that a single transit measurement
consists of observing the primary eclipse and an equal out-of-transit
baseline \citep[cf.][]{kreidberg2014a}.
We check our transit estimate by comparing our SNR
with the detailed calculations in \citet{batalha2015a}, and find
that we can reproduce their results up to a factor of two (not shown).
We also note that our estimate of GJ 1132b's thermal emission is
slightly higher than the signal in \citet{berta-thompson2015}, because
the observer-projected dayside temperature of a bare rock is higher
than its equilibrium temperature by $(8/3)^{1/4}\approx 1.28$
\citep{koll2015}.

%%%
\begin{table}[t!]
\tablenum{2}
\begin{center}
    \begin{tabular}{ l l c c c }
%% \tableline
      %\toprule
    \multicolumn{5}{c}{\textbf{Transit vs.~phase curve observations of
      GJ 1132b with \textit{JWST}}}\\
    % \addlinespace
      Method & Observation Time & Signal (ppm) &Noise (ppm) & SNR \\
\tableline
    % \toprule
    % \addlinespace
     Single transit\tablenotemark{a} & one transit\tablenotemark{b} = 90 min& 19.9 & 19.7 & 1\\
    % \addlinespace
     Stacked transits\tablenotemark{a} & 13 transits = 19.5 hours& 19.9 & 5.5 & 4 \\
    % \addlinespace
    Thermal phase curve\tablenotemark{c}& one half-orbit = 19.5 hours& 373 & 84 & 4 \\
    % \bottomrule
\tableline
    \end{tabular}\\
\end{center}
    \tablenotetext{a}{$1-4\mu$m, NIRSpec, $R\sim25$, CO$_2$-dominated atmosphere.}
    \tablenotetext{b}{We assume a measurement lasts $45$ min in-transit, plus
      $45$ min out-of-transit baseline.}
    \tablenotetext{c}{$16.5-19.5\mu$m, MIRI, broadband.}
\caption{Transit spectroscopy and thermal phase curve measurements of a planet
  like GJ 1132b will require similar amounts
  of \textit{JWST} observation time. The shown signal-to-noise
  (SNR) ratios are estimates for the most basic
  observational goals: detecting molecular features in
  low-resolution near-IR transit spectra, and detecting
  the day-night thermal emission contrast of a bare
  rock in the mid-IR. We compute signals following \citet{cowan2015} and
  \citet{koll2015}. We estimate noise
  assuming photon-limited precision, but include imperfect
  instrument throughput for MIRI (see Section \ref{sec:obs}).
  \label{tab:feasibility}}
\end{table}
%%%

%
Table \ref{tab:feasibility} shows our results. Similar to previous
estimates \citep[e.g.,][]{batalha2015a,cowan2015}, we find that a
single transit would not be sufficient to conclusively identify
molecular absorption features (SNR $\sim1$). The low SNR arises
largely because of the high MMW atmosphere; for comparison, a
H$_2$ atmosphere on GJ 1132b should be detectable in a single
transit with SNR $\sim22$. For a CO$_2$ atmosphere,
13 repeated transit observations would reduce the noise sufficiently
to allow spectral features to be discerned with SNR $\sim 4$. The time
it takes to measure 13 repeated transits of GJ 1132b is also equal to
the time it takes to measure one half-orbit phase curve (from transit
to secondary eclipse). We find that the thermal emission of a bare
rock would be detectable with a comparable SNR $\sim 4$.
We conclude that characterizing high MMW atmospheres of rocky
exoplanets will require relatively large investments of \textit{JWST}
time. If such observations are pursued, however, then thermal phase
curves are a feasible technique that would yield important
complementary information about these planets \citep{koll2015}.

\section{Discussion}
\label{sec:discussion}

The heat engine framework is well-established for Earth's atmosphere
\citep[e.g.,][]{peixoto1984}. Similarly \citet{goodman2009} pointed
out that hot Jupiters can be viewed as heat engines, but did not
develop his insight more quantitatively. Here we have
demonstrated that the atmospheres of rocky exoplanets act as heat
engines, which allowed us to develop a new constraint on their
day-night circulations.
We also found that surface wind speeds in most of our GCM simulations
are about a factor of two smaller than the value predicted by the heat
engine (Fig.~\ref{fig:Usurf1}). Because work scales with the cube of the
surface wind speed, our simulations produce $\sim(1/2)^{3}=1/8$ as
much work as an ideal heat engine. Interestingly, Earth's atmospheric heat
engine also produces about an order of magnitude less work than its
ideal limit and therefore has a similar
inefficiency as our dry and tidally locked simulations \citep{peixoto1984}.
Our result seems to be at odds with the usual understanding that the inefficiency of
Earth's atmospheric heat engine is caused by its hydrological cycle
\citep{pauluis2000, pauluis2010}, and also raises the question whether our
scaling can be generalized to planets that are not tidally locked. We hope to
address these issues in future work.
Our results also strongly suggest that hot Jupiters should obey
similar constraints as rocky planets. For example, using the heat
engine framework it might be possible to constrain the day-night
overturning circulation, which controls the vertical mixing and
chemical equilibrium of hot Jupiter atmospheres.
However, modeling these atmospheres as heat engines
will require a better understanding of the mechanisms through which they
dissipate kinetic energy, which could include magneto-hydrodynamic
drag, shocks, or shear instabilities \citep{li2010,menou2012b,fromang2016}.

Our results allow us to interpret previous GCM results that have not been
fully explained yet.
First, \citet{merlis2010} explored Earth-like atmospheres at different
rotation rates. They found that although the strength of superrotation
is strongly dependent on rotation rate, day-night surface temperature
gradients are mostly insensitive to rotation rate (their Fig.~15). Our
results explain why: \citeauthor{merlis2010} varied rotation rates
while keeping the stellar flux fixed at Earth's value. Their
simulations were therefore in the rapidly rotating ($a^2/L_{Ro}^2>1$)
but relatively cool/thick regime ($t_{wave}/t_{rad} < 10^{-2}$) in
which temperature structure is not strongly sensitive to rotation
(Fig.~\ref{fig:Ttrop1}).
Second, in \citet{koll2015} we
found that thermal phase curves are mainly sensitive
to the nondimensional parameters $t_{wave}/t_{rad}$ and
$\tau_{LW}$. Our result only broke down for hot/thin and rapidly
rotating atmospheres, $t_{wave}/t_{rad}\gtrsim 10^{-2}$ and
$a^2/L_{Ro}^2\gtrsim1$. Our results here explain both the
wave-to-radiative timescale threshold of $t_{wave}/t_{rad}\sim10^{-2}$
and why rotation is relatively unimportant (Figs.~\ref{fig:dTemp1}
and \ref{fig:Ttrop1}).

There are additional physical effects that might affect our conclusions.
We assume broadband grey radiative transfer, but a wide range of
plausible atmospheric compositions feature significant spectral window
regions. As already noted by \citet{leconte2013} and
\citet{wordsworth2015}, window regions allow the nightside surface to
cool even more effectively, which would increase day-night surface
temperature gradients compared to that predicted by our grey models.
At the same time, large spectral window regions would increase the
atmosphere's radiative cooling timescale and thus reduce atmospheric
temperature gradients, similar to the optically thin cases we
discussed in Section \ref{sec:transition}.
We also did not consider shortwave absorption. Shortwave
absorption will shift heating to lower pressures, which would
decrease the heat intake temperature of the atmospheric heat
engine and reduce the atmospheric circulation strength. We
therefore expect our heat engine theory to be an upper bound on wind
speeds.

Many planets might be able to retain a hydrologic cycle (e.g., H$_2$O
inside the habitable zone, or CH$_4$ on Titan-like planets) against
atmospheric escape and nightside collapse. Besides changing the
atmosphere's radiative properties (e.g., H$_2$O effectively absorbs
both in the shortwave and longwave), condensation would also modify
the atmospheric dynamics.
Moist GCM simulations indicate that the temperature and circulation
structure sketched out in Figure \ref{fig:diagRadSub} could still
apply qualitatively \citep{merlis2010,yang2013}, but with several
modifications. First, latent heat transport would reduce the day-night
temperature gradient compared to dry atmospheres
\citep{leconte2013}. Second, moist convection would lead to thick
cloud cover on the dayside and could drastically change a planet's
appearance to remote observers \citep{fortney2005,yang2013}.  Third,
dry atmospheres develop strongly turbulent daysides. The friction
associated with this dry convection allows the nightside temperature
structure to decouple from regions of convection
\citep[Fig.~\ref{fig:climDry50a},][]{voigt2012}. In contrast, moist
atmospheres such as Earth's tropics maintain an adiabatic temperature
profile through deep moist convection, while the dry turbulent
boundary layer is relatively shallow. We therefore expect that moist
atmospheres would be less dominated by friction, and would be even
better captured by WTG models similar to our radiative-convective
model (Section \ref{sec:radconv}).

It is an open question how many rocky planets around M-stars will
actually be tidally locked.  \citet{leconte2015} found that thermal
tides in relatively thick atmospheres ($p_s \gtrsim 1$ bar) can
prevent habitable-zone planets around early M-dwarfs from reaching a
tidally locked state. Although thermal tides could limit the application of our
results to planets on longer-period orbits, they are less likely to apply
to planets around late M-dwarfs or hot exoplanets like GJ
1132b. Moreover, given that rocky exoplanets are extremely common, we
also expect that future discoveries will find rocky exoplanets in a
wide range of rotational states. Optical phase
curves could constrain the rotation rates of these planets
without relying on models \citep{fujii2014}, while future theoretical
work should consider the connection between the tidally locked limit
we considered here and planets in higher-order spin-orbit resonances.

\section{Conclusions}
\label{sec:conclusions}

We have developed a series of theoretical models to understand the basic
temperature structure and large-scale circulations of tidally locked
planets with dry atmospheres. These models are able to capture and
predict many fundamental aspects of much more complex GCM simulations,
including the atmospheric temperature structure, dayside and nightside
surface temperatures, as well as large-scale wind speeds.
We draw the following conclusions from our work:

\begin{enumerate}
\item Our radiative-convective model describes tidally locked
  atmospheres with efficient day-night heat transport and applies in
  the limit of cool and thick atmospheres
  ($t_{wave}/t_{rad} \lesssim10^{-4}$).  It captures the basic
  temperature structure of tidally locked planets and extends the
  asymptotic theory for optically thick atmospheres
  \citep[$\tau_{LW}\gg1$,][]{pierrehumbert2011b} to arbitrary optical
  thickness.

\item Atmospheres of dry, tidally locked exoplanets act as global heat
  engines. Our heat engine scaling
  places strong constraints on the day-night circulation strength of
  tidally locked atmospheres.

\item Our radiative-convective-subsiding model describes tidally
  locked atmospheres with limited day-night heat transport. It extends
  both our radiative-convective model and the asymptotic theory for
  optically thin atmospheres
  \citep[$\tau_{LW}\ll1$,][]{wordsworth2015}, and captures the dynamics
  of a wide range of complex GCM simulations.
  It breaks down in the limit of
  atmospheres that are both rapidly rotating ($a^2/L_{Ro}^2 \gtrsim
  1$) and hot/thin ($t_{wave}/t_{rad} > \mathcal{O}(10^{-2})$).

\item Like hot Jupiters, day-night atmospheric temperature
  gradients of rocky exoplanets become large once parcels of
  air take longer to subside than to cool radiatively.
  Unlike hot Jupiters, the timescale for subsidence on rocky
  planets is severely increased by the limited heat engine efficiency
  and the areal asymmetry between convection and subsidence.
  Rocky planets develop large day-night atmospheric temperature gradients when 
  \begin{eqnarray}
    \frac{t_{wave}}{t_{rad}} & \gtrsim & 
                                         \begin{cases} \chi^{3/2} \times \dfrac{c_p}{R}
                                           \left( \dfrac{t_{drag}}{t_{wave}} \right)^{1/2}&
                                           ~\text{if } \tau_{LW} \geq 1,\\ 
                                           \dfrac{\chi^{3/2}}{\tau_{LW}} \times \dfrac{c_p}{R}
                                           \left(
                                             \dfrac{t_{drag}}{t_{wave}} \right)^{1/2} &
                                           ~\text{if } \tau_{LW} < 1.
                                         \end{cases}
  \end{eqnarray}
  Optically thin atmospheres cool inefficiently, which makes
  them less likely to develop large temperature gradients than
  optically thick atmospheres.

\item Rapid rotation ($a^2/L_{Ro}^2 \gtrsim 1$) only has a strong
  influence on temperature structure if the wave-to-radiative
  timescale exceeds the above ratio, $t_{wave}/t_{rad} \gtrsim \mathcal{O}(10^{-2})$.
  Once rotation is important its effects cannot be ignored for
  a detailed understanding of a planet's atmosphere, including its
  thermal phase curve signature and the potential for atmospheric
  collapse.

\item Short-period rocky exoplanets with high MMW atmospheres and
  surface pressures of $\lesssim 1$ bar will likely exhibit significant
  day-night temperature gradients. Thermal phase curve
  observations of such planets will require
  similar amounts of \textit{JWST} time as transit observations.

\end{enumerate}

\acknowledgments

Our work was completed with resources provided by the University of
Chicago Research Computing Center. We thank Malte Jansen and Feng Ding for
insightful discussions, and an anonymous reviewer for
their feedback. D.D.B.~Koll was supported by a William Rainey
Harper dissertation fellowship.

\clearpage

\appendix
\section{Appendix: Drag Timescale}
\label{sec:appendix1}
We start with the nondimensional parameters that we derived in
\citet{koll2015}:
\begin{eqnarray}
  \left( \frac{R}{c_p}, \frac{a^2}{L_{Ro}^2},
    \frac{t_{wave}}{t_{rad}},\tau_{SW},\tau_{LW}, \frac{C_D a}{H} \right).
\end{eqnarray}
Theoretical work on the atmospheric dynamics of hot
Jupiters uses similar wave and
radiative timescales \citep[e.g.,][]{perez-becker2013a}, but additionally introduces a
drag timescale. This is because the drag mechanisms on
hot Jupiters are still not well-constrained, and friction is often
parametrized as Rayleigh friction with a unknown damping timescale.
To facilitate comparison of our work with the hot Jupiter literature
and to examine the importance of drag in the atmospheres of rocky planets,
we rewrite the last of our six nondimensional
parameters as a ratio of wave over drag timescales.

Models of terrestrial atmospheres (including FMS) often parametrize boundary layer
friction as vertical momentum diffusion with a source term that is quadratic in wind speed. The
horizontal momentum equation takes the form
\begin{eqnarray}
  \DDt{\mathbf{u}} &=& \dots + g \frac{\partial \bm{\mathcal{D}}_m}{\partial p},
\end{eqnarray}
where $\mathbf{u}$ is the horizontal wind speed, $\bm{\mathcal{D}}_m$ is the
diffusive momentum flux due to surface drag, and the source term is
$\bm{\mathcal{D}}_m\left(p_s\right) = C_D \rho_s |\mathbf{u}_s| \mathbf{u}_s$.  We
take a vertical average across the boundary layer to find the average
acceleration due to drag:
\begin{eqnarray}
  \frac{1}{p_s-p_{BL}} \int_{p_{BL}}^{p_s} \DDt{\mathbf{u}} dp & = & \dots
  + \frac{g}{p_s-p_{BL}} \int_{p_{BL}}^{p_s} \frac{\partial
    \bm{\mathcal{D}}_m}{\partial p} dp \nonumber\\
\overline{\DDt{\mathbf{u}}} & = & \dots
  + \frac{g}{p_s-p_{BL}} \times
                                        \left[ \bm{\mathcal{D}}_m\left(p_s\right) -
                                        \bm{\mathcal{D}}_m\left(p_{BL}\right)
                                        \right] \nonumber\\
\overline{\DDt{\mathbf{u}}} & = & \dots
  + \frac{g \bm{\mathcal{D}}_m\left(p_s\right)}{p_s-p_{BL}}\nonumber\\
\overline{\DDt{\mathbf{u}}}& = & \dots
  + \frac{g C_D \rho_s |\mathbf{u}_s| \mathbf{u}_s}{p_s-p_{BL}}
\end{eqnarray}
Here $p_{BL}$ denotes the top pressure level of the boundary layer and
we used the fact that drag has to disappear at the upper edge of
the boundary layer, $\bm{\mathcal{D}}_m\left(p_{BL}\right)=0$.
We then scale this
equation for fast atmospheric motions $\mathbf{u}\sim
c_{wave}$,
\begin{eqnarray}
  \frac{c_{wave}}{t_{drag}} & \sim & \frac{g C_D \rho_s
                                     c_{wave}^2}{p_s} \nonumber\\
  \frac{c_{wave}}{t_{drag}} & \sim & \frac{a C_D g}{R T_s}
                                     \frac{c_{wave}}{a} c_{wave}\nonumber \\
  \frac{c_{wave}}{t_{drag}} & \sim & \frac{C_D a}{H}
                                     \frac{1}{t_{wave}} c_{wave},
\end{eqnarray}
where we have assumed that the boundary layer is thick, $p_{BL} \ll p_s$ (see Fig.~\ref{fig:climDry50a}), used the ideal gas law in the second step,
$\rho_s = p_s R^{-1} T_s^{-1}$, and used the wave timescale $t_{wave}
= a/c_{wave}$ in the last step. This lets us derive a drag timescale
\begin{eqnarray}
  t_{drag} & \sim & \frac{H}{C_D a} t_{wave}.
\end{eqnarray}
Note, in contrast to Rayleigh drag schemes where the drag
timescale is independent of $\mathbf{u}$, here the drag timescale scales
with $\mathbf{u}$ and thus with the dynamical timescale $t_{wave}$.
Using this drag timescale we can rewrite the last nondimensional
parameter as $C_D a/H \sim t_{wave}/t_{drag}$ and find an alternative
set of six governing parameters:
\begin{eqnarray}
  \left( \frac{R}{c_p}, \frac{a^2}{L_{Ro}^2},
    \frac{t_{wave}}{t_{rad}},\tau_{SW},\tau_{LW}, \frac{t_{wave}}{t_{drag}} \right).
\end{eqnarray}

In most cases $C_D a/H$ is of order unity so the drag timescale is
generally comparable to the dynamical timescale.  For example,
assuming a planet of Earth's size, $(a,g)=(a_\Earth,g_{\Earth})$, a
high MMW atmosphere, $R=R_{N_2}$, a relatively cool
temperature, $T_{eq}=300K$, and a standard value for the drag
coefficient, $C_D=10^{-3}$, we find $t_{drag} = 1.4 t_{wave}$.
Variations in the planetary radius $a$ or in the drag coefficient
$C_D$ do not affect this result much. For example, for a neutrally buoyant
boundary layer $C_D = [k_{vk}/\log(z/z_0)]^2$, where $k_{kv}$ is the
von Karman constant, $z$ is the height above the surface and $z_0$ is
the surface roughness length. Because $C_D$ only depends
logarithmically on $z_0$, the drag timescale is not very sensitive to
the surface properties.
This means we expect friction to generally be
an important process inside the boundary layer of rocky planets.

The most important exception is a hot H$_2$ atmosphere, through its
effect on the scale height $H$. For example, repeating the above
calculation with a hot H$_2$-dominated
atmosphere, $R=R_{H_2}$ and $T_{eq}=600$ K, we find $t_{drag} = 40
t_{wave}$. This means surface friction is
far less effective in H$_2$ atmospheres than in high
MMW atmospheres.

\section{Appendix: Wind speed scaling from Wordsworth (2015)}
\label{sec:appendix2}

To compare the results of \citet{wordsworth2015} with our
GCM simulations we write \citeauthor{wordsworth2015}'s Equation 33
as
\begin{eqnarray}
  U_0 & = & 4 \sigma T_{eq}^4 \frac{\tau_{LW}}{2 \zeta p_s C_D} \frac{R}{c_p},
\end{eqnarray}
where $\zeta \equiv 1/3$. The above equation reduces to
\citeauthor{wordsworth2015}'s Equation 33 by plugging in his Equation
12, i.e., by assuming a specific form for $\tau$. We leave the
equation in this general form. We identify
\citeauthor{wordsworth2015}'s absorbed stellar flux $(1-A)F$ as $4\sigma T_{eq}^4$.

The nondimensional equations in \citet{wordsworth2015} are not
affected. To find the surface wind speed we solve his Equations 44
and 45 numerically to find $\tilde{T}$ and $\tilde{U}$. We then
convert the nondimensional $\tilde{U}$ into a dimensional quantity
using the above scale $U_0$, i.e., $|\mathbf{u}| = U_0 \tilde{U}$.

%%%% -----
\section{Appendix: Numerical solution for the RCS model}
\label{sec:appendix3}

The boundary conditions of the
  radiative-convective-subsiding (RCS) model are specified at two
  different points, the tropopause and the surface. Instead of
  matching both boundaries simultaneously we first guess a value of
  the nightside OLR, $F(\tau_0)$. Given a value of $F(\tau_0)$, we can
  solve for all variables ($T_d$, $\tau_0$, $T(\tau)$, and
  $F(\tau)$). Our guess will in general not satisfy the nightside
  surface energy budget (Equation \ref{eqn:rcs4}c), so we iterate
  until $F(\tau_{LW})=0$ is satisfied.  To iterate we use a bisection
  method, where the nightside OLR is bounded by
  $0\leq F(\tau_0) \leq \sigma T_{eq}^4$ (the limits correspond to a
  planet with zero and perfect day-night heat redistribution).

We proceed as follows. Given a value of
  $F(\tau_0)$, Equations \ref{eqn:rcs5} and \ref{eqn:rcs3} can be
  rewritten as an implicit equation for $\tau_0$,
\begin{eqnarray}
  2+\frac{\tau_0}{2} & = &
        \left(\frac{1+\tau_0}{2}\right)
                                        \left(\frac{\tau_{LW}}{\tau_{0}}\right)^{4\beta} \left[ e^{-(\tau_{LW}-\tau_0)}
         + \int_{\tau_0}^{\tau_{LW}} \left(\frac{\tau'}{\tau_{LW}}\right)^{4\beta}
          e^{-(\tau'-\tau_0)} d\tau' \right] + \frac{F(\tau_0)}{\sigma T_{eq}^4}
\end{eqnarray}
We solve for $\tau_0$ using, again, a bisection method, where
  we note that $0<\tau_0<\tau_{LW}$. We then use Equations \ref{eqn:rcs2}
  and \ref{eqn:rcs3} to find $T(\tau_0)$ and $T_d$:
\begin{eqnarray}
  T(\tau_0) & = & T_{eq} \left( \frac{1+\tau_0}{2} \right)^{1/4} \\
  T_d & = & T_{eq} \left( \frac{1+\tau_0}{2} \right)^{1/4} \left(
            \frac{\tau_{LW}}{\tau_0} \right)^\beta.
\end{eqnarray}
Once we know $T_d$ we can find
  $\bar{\omega}$ (Equations \ref{eqn:he1} and \ref{eqn:rcs6}). We then
  have three boundary conditions that are specified at the upper
  boundary, the guessed $F(\tau_0)$, $dF(\tau_0)/d\tau=0$, and
  $T(\tau_0)$. We use SciPy's VODE solver
  to integrate the WTG and Schwarzschild equations (Equations
  \ref{eqn:rcs1}) down to the nightside surface, which gives us the
  net surface flux $F(\tau_{LW})$. We iterate until we
  satisfy the nightside surface budget, $F(\tau_{LW})=0$. After having
  solved for $T(\tau)$ we find the
  nightside surface temperature $T_n$ using the nightside surface
  energy budget \citep[cf.][]{robinson2012},
\begin{eqnarray}
  \sigma T_n^4 & = & F^-(\tau_{LW}) \nonumber \\
    & = & \sigma T_{eq}^4 \frac{\tau_0}{2} e^{-(\tau_{LW}-\tau_0)} +
                     \int_{\tau_0}^{\tau_{LW}} \sigma T(\tau')^4 e^{-(\tau'-\tau_0)} d\tau'.
\end{eqnarray}

%%%% -----

\section{Appendix: Equations of Motion}
\label{sec:appendix4}

The FMS GCM integrates the primitive equations in pressure coordinates, which are
\begin{eqnarray}
\frac{D\mathbf{u}}{Dt} &=& -f \mathbf{k} \times \mathbf{u} - \nabla\phi -g \frac{\partial \bm{\mathcal{D}}_{m} }{\partial p}, \label{eqn:consmom}\\
\frac{\partial \phi}{\partial p}&=& -\frac{R T}{p}, \label{eqn:hydro}\\
\nabla \cdot \mathbf{u} & + & \frac{\partial \omega}{\partial p}= 0,
                              \label{eqn:consmass} \\
\frac{DT}{Dt}&=& \frac{R T \omega}{c_p p} + \frac{g}{c_p}
\frac{\partial F}{\partial p} + \frac{g}{c_p}
\frac{\partial \mathcal{D}}{\partial p}. \label{eqn:consenergy}
\end{eqnarray}
Here $\mathbf{u} = (u,v)$ is the horizontal wind velocity,
$\frac{D}{Dt} = \frac{\partial}{\partial t} + \mathbf{u} \cdot \nabla
+ \omega \frac{\partial}{\partial p}$
is the material derivative, $f=2 \Omega \sin \theta$ is the Coriolis
parameter, $\mathbf{k}$ is local vertical unit vector, $\phi$ is the
geopotential, $T$ is temperature, $\omega \equiv Dp/Dt$ is the
pressure velocity, $\bm{\mathcal{D}}_m$ and $\mathcal{D}$ are the
vertical diffusive fluxes of momentum and energy in the boundary
layer, $F$ is the net longwave flux, and an overview of the
dimensional parameters can be found in Table \ref{tab:minmax}. From
the top, these equations express conservation of momentum, the
hydrostatic approximation, conservation of mass, and conservation of
energy.  The net longwave flux $F$ is governed by the two-stream
Schwarzschild equations. For a grey gas these can be written in
optical depth coordinates as
\begin{eqnarray}
\frac{\partial^2 F }{\partial \tau^2} - F &=& - 2 \frac{\partial
                                              (\sigma T^4)}{\partial \tau}. \label{eqn:twostream}
\end{eqnarray}

\clearpage

\bibliography{ZoteroLibrary_short}

\begin{thebibliography}{}
\expandafter\ifx\csname natexlab\endcsname\relax\def\natexlab#1{#1}\fi

\bibitem[{Abe {et~al.}(2011)Abe, Abe-Ouchi, Sleep, \& Zahnle}]{abe2011}
Abe, Y., Abe-Ouchi, A., Sleep, N.~H., \& Zahnle, K.~J. 2011, Astrobiology, 11,
  443

\bibitem[{Batalha {et~al.}(2015)Batalha, Kalirai, Lunine, Clampin, \&
  Lindler}]{batalha2015a}
Batalha, N., Kalirai, J., Lunine, J., Clampin, M., \& Lindler, D. 2015,
  arXiv:1507.02655 [astro-ph], arXiv:1507.02655

\bibitem[{Beichman {et~al.}(2014)Beichman, Benneke, Knutson, Smith, Lagage,
  Dressing, Latham, Lunine, Birkmann, Ferruit, Giardino, Kempton, Carey, Krick,
  Deroo, Mandell, Ressler, Shporer, Swain, Vasisht, Ricker, Bouwman,
  Crossfield, Greene, Howell, Christiansen, Ciardi, Clampin, Greenhouse,
  Sozzetti, Goudfrooij, Hines, Keyes, Lee, McCullough, Robberto, Stansberry,
  Valenti, Rieke, Rieke, Fortney, Bean, Kreidberg, Ehrenreich, Deming, Albert,
  Doyon, \& Sing}]{beichman2014}
Beichman, C., Benneke, B., Knutson, H., {et~al.} 2014, Publications of the
  Astronomical Society of the Pacific, 126, 1134

\bibitem[{Berta-Thompson {et~al.}(2015)Berta-Thompson, Irwin, Charbonneau,
  Newton, Dittmann, Astudillo-Defru, Bonfils, Gillon, Jehin, Stark, Stalder,
  Bouchy, Delfosse, Forveille, Lovis, Mayor, Neves, Pepe, Santos, Udry, \&
  W{\"u}nsche}]{berta-thompson2015}
Berta-Thompson, Z.~K., Irwin, J., Charbonneau, D., {et~al.} 2015, Nature, 527,
  204

\bibitem[{Bister \& Emanuel(1998)}]{bister1998}
Bister, M., \& Emanuel, K.~A. 1998, Meteorology and Atmospheric Physics, 65,
  233

\bibitem[{Buckingham(1914)}]{buckingham1914}
Buckingham, E. 1914, Physical Review, 4, 345

\bibitem[{Charnay {et~al.}(2015)Charnay, Meadows, Misra, Leconte, \&
  Arney}]{charnay2015b}
Charnay, B., Meadows, V., Misra, A., Leconte, J., \& Arney, G. 2015, The
  Astrophysical Journal Letters, 813, L1

\bibitem[{Cowan {et~al.}(2015)Cowan, Greene, Angerhausen, Batalha, Clampin,
  Col{\'o}n, Crossfield, Fortney, Gaudi, Harrington, Iro, Lillie, Linsky,
  Lopez-Morales, Mandell, Stevenson, \& SAG-10}]{cowan2015}
Cowan, N.~B., Greene, T., Angerhausen, D., {et~al.} 2015, Publications of the
  Astronomical Society of the Pacific, 127, 311

\bibitem[{Deming {et~al.}(2009)Deming, Seager, Winn, Miller-Ricci, Clampin,
  Lindler, Greene, Charbonneau, Laughlin, Ricker, Latham, \&
  Ennico}]{deming2009c}
Deming, D., Seager, S., Winn, J., {et~al.} 2009, Publications of the
  Astronomical Society of the Pacific, 121, 952

\bibitem[{Dressing \& Charbonneau(2015)}]{dressing2015}
Dressing, C.~D., \& Charbonneau, D. 2015, The Astrophysical Journal, 807, 45

\bibitem[{Emanuel(1986)}]{emanuel1986}
Emanuel, K.~A. 1986, Journal of the Atmospheric Sciences, 43, 585

\bibitem[{Emanuel \& Bister(1996)}]{emanuel1996}
Emanuel, K.~A., \& Bister, M. 1996, Journal of the Atmospheric Sciences, 53,
  3276

\bibitem[{Emanuel \& Rotunno(1989)}]{emanuel1989}
Emanuel, K.~A., \& Rotunno, R. 1989, Tellus A, 41,
  doi:10.3402/tellusa.v41i1.11817

\bibitem[{Fortney(2005)}]{fortney2005}
Fortney, J.~J. 2005, Monthly Notices of the Royal Astronomical Society, 364,
  649

\bibitem[{Frierson {et~al.}(2006)Frierson, Held, \&
  Zurita-Gotor}]{frierson2006}
Frierson, D. M.~W., Held, I.~M., \& Zurita-Gotor, P. 2006, Journal of the
  Atmospheric Sciences, 63, 2548

\bibitem[{Fromang {et~al.}(2016)Fromang, Leconte, \& Heng}]{fromang2016}
Fromang, S., Leconte, J., \& Heng, K. 2016, Astronomy \& Astrophysics,
  doi:10.1051/0004-6361/201527600

\bibitem[{Fujii {et~al.}(2014)Fujii, Kimura, Dohm, \& Ohtake}]{fujii2014}
Fujii, Y., Kimura, J., Dohm, J., \& Ohtake, M. 2014, Astrobiology, 14, 753

\bibitem[{Glasse {et~al.}(2010)Glasse, Bauwens, Bouwman, Detre, Fischer,
  Garcia-Marin, Justannont, Labiano, Nakos, Ressler, Rieke, Scheithauer, Wells,
  \& Wright}]{glasse2010}
Glasse, A. C.~H., Bauwens, E., Bouwman, J., {et~al.} 2010, in Proc. {{SPIE}}
  7731, ed. J.~M. {Oschmann, Jr.}, M.~C. Clampin, \& H.~A. MacEwen, 77310K

\bibitem[{Goodman(2009)}]{goodman2009}
Goodman, J. 2009, The Astrophysical Journal, 693, 1645

\bibitem[{Heng \& Kopparla(2012)}]{heng2012a}
Heng, K., \& Kopparla, P. 2012, The Astrophysical Journal, 754, 60

\bibitem[{Heng {et~al.}(2011)Heng, Menou, \& Phillipps}]{heng2011a}
Heng, K., Menou, K., \& Phillipps, P.~J. 2011, Monthly Notices of the Royal
  Astronomical Society, 413, 2380

\bibitem[{Hu {et~al.}(2015)Hu, Demory, Seager, Lewis, \& Showman}]{hu2015}
Hu, R., Demory, B.-O., Seager, S., Lewis, N., \& Showman, A.~P. 2015, The
  Astrophysical Journal, 802, 51

\bibitem[{Joshi {et~al.}(1997)Joshi, Haberle, \& Reynolds}]{joshi1997}
Joshi, M., Haberle, R., \& Reynolds, R. 1997, Icarus, 129, 450

\bibitem[{Kaltenegger \& Traub(2009)}]{kaltenegger2009}
Kaltenegger, L., \& Traub, W.~A. 2009, The Astrophysical Journal, 698, 519

\bibitem[{Kaspi \& Showman(2015)}]{kaspi2015}
Kaspi, Y., \& Showman, A.~P. 2015, The Astrophysical Journal, 804, 60

\bibitem[{Kasting(1988)}]{kasting1988}
Kasting, J.~F. 1988, Icarus, 74, 472

\bibitem[{Kasting {et~al.}(1993)Kasting, Whitmire, \& Reynolds}]{kasting1993a}
Kasting, J.~F., Whitmire, D.~P., \& Reynolds, R.~T. 1993, Icarus, 101, 108

\bibitem[{Koll \& Abbot(2015)}]{koll2015}
Koll, D. D.~B., \& Abbot, D.~S. 2015, The Astrophysical Journal, 802, 21

\bibitem[{Komacek \& Showman(2016)}]{komacek2016}
Komacek, T.~D., \& Showman, A.~P. 2016, The Astrophysical Journal, 821, 16

\bibitem[{Kopparapu {et~al.}(2016)Kopparapu, Wolf, Haqq-Misra, Yang, Kasting,
  Meadows, Terrien, \& Mahadevan}]{kopparapu2016}
Kopparapu, R.~K., Wolf, E.~T., Haqq-Misra, J., {et~al.} 2016, The Astrophysical
  Journal, 819, 84

\bibitem[{Kreidberg {et~al.}(2014)Kreidberg, Bean, D{\'e}sert, Benneke, Deming,
  Stevenson, Seager, Berta-Thompson, Seifahrt, \& Homeier}]{kreidberg2014a}
Kreidberg, L., Bean, J.~L., D{\'e}sert, J.-M., {et~al.} 2014, Nature, 505, 69

\bibitem[{Leconte {et~al.}(2013)Leconte, Forget, Charnay, Wordsworth, Selsis,
  Millour, \& Spiga}]{leconte2013}
Leconte, J., Forget, F., Charnay, B., {et~al.} 2013, Astronomy \& Astrophysics,
  554, A69

\bibitem[{Leconte {et~al.}(2015)Leconte, Wu, Menou, \& Murray}]{leconte2015}
Leconte, J., Wu, H., Menou, K., \& Murray, N. 2015, Science, 347, 632

\bibitem[{Li \& Goodman(2010)}]{li2010}
Li, J., \& Goodman, J. 2010, The Astrophysical Journal, 725, 1146

\bibitem[{Line \& Parmentier(2016)}]{line2016}
Line, M.~R., \& Parmentier, V. 2016, The Astrophysical Journal, 820, 78

\bibitem[{Liu \& Schneider(2011)}]{liu2011}
Liu, J., \& Schneider, T. 2011, Journal of the Atmospheric Sciences, 68, 2742

\bibitem[{Makarov {et~al.}(2012)Makarov, Berghea, \& Efroimsky}]{makarov2012a}
Makarov, V.~V., Berghea, C., \& Efroimsky, M. 2012, The Astrophysical Journal,
  761, 83

\bibitem[{Matsuno(1966)}]{matsuno1966}
Matsuno, T. 1966, Journal of the Meteorological Society of Japan. Ser. II, 44,
  25

\bibitem[{Menou(2012{\natexlab{a}})}]{menou2012a}
Menou, K. 2012{\natexlab{a}}, The Astrophysical Journal Letters, 744, L16

\bibitem[{Menou(2012{\natexlab{b}})}]{menou2012b}
---. 2012{\natexlab{b}}, The Astrophysical Journal, 745, 138

\bibitem[{Merlis \& Schneider(2010)}]{merlis2010}
Merlis, T.~M., \& Schneider, T. 2010, Journal of Advances in Modeling Earth
  Systems, 2, doi:10.3894/JAMES.2010.2.13

\bibitem[{Mills \& Abbot(2013)}]{mills2013}
Mills, S.~M., \& Abbot, D.~S. 2013, The Astrophysical Journal Letters, 774, L17

\bibitem[{Motalebi {et~al.}(2015)Motalebi, Udry, Gillon, Lovis, S{\'e}gransan,
  Buchhave, Demory, Malavolta, Dressing, Sasselov, Rice, Charbonneau, {Collier
  Cameron}, Latham, Molinari, Pepe, Affer, Bonomo, Cosentino, Dumusque,
  Figueira, Fiorenzano, Gettel, Harutyunyan, Haywood, Johnson, Lopez,
  Lopez-Morales, Mayor, Micela, Mortier, Nascimbeni, Philips, Piotto, Pollacco,
  Queloz, Sozzetti, Vanderburg, \& Watson}]{motalebi2015}
Motalebi, F., Udry, S., Gillon, M., {et~al.} 2015, Astronomy \& Astrophysics,
  584, A72

\bibitem[{Owen \& Mohanty(2016)}]{owen2016}
Owen, J.~E., \& Mohanty, S. 2016, Monthly Notices of the Royal Astronomical
  Society, stw959

\bibitem[{Parmentier {et~al.}(2013)Parmentier, Showman, \&
  Lian}]{parmentier2013}
Parmentier, V., Showman, A.~P., \& Lian, Y. 2013, Astronomy \& Astrophysics,
  558, A91

\bibitem[{Pauluis(2010)}]{pauluis2010}
Pauluis, O. 2010, Journal of the Atmospheric Sciences, 68, 91

\bibitem[{Pauluis {et~al.}(2000)Pauluis, Balaji, \& Held}]{pauluis2000}
Pauluis, O., Balaji, V., \& Held, I.~M. 2000, Journal of the Atmospheric
  Sciences, 57, 989

\bibitem[{Pauluis \& Held(2002)}]{pauluis2002a}
Pauluis, O., \& Held, I.~M. 2002, Journal of the Atmospheric Sciences, 59, 125

\bibitem[{Peixoto \& Oort(1984)}]{peixoto1984}
Peixoto, J.~P., \& Oort, A.~H. 1984, Reviews of Modern Physics, 56, 365

\bibitem[{Perez-Becker \& Showman(2013)}]{perez-becker2013a}
Perez-Becker, D., \& Showman, A.~P. 2013, The Astrophysical Journal, 776, 134

\bibitem[{Pierrehumbert(2011{\natexlab{a}})}]{pierrehumbert2011c}
Pierrehumbert, R.~T. 2011{\natexlab{a}}, The Astrophysical Journal, 726, L8

\bibitem[{Pierrehumbert(2011{\natexlab{b}})}]{pierrehumbert2011b}
---. 2011{\natexlab{b}}, Principles of {{Planetary Climate}} ({Cambridge
  University Press})

\bibitem[{Renno \& Ingersoll(1996)}]{renno1996}
Renno, N.~O., \& Ingersoll, A.~P. 1996, Journal of the Atmospheric Sciences,
  53, 572

\bibitem[{Robinson \& Catling(2012)}]{robinson2012}
Robinson, T.~D., \& Catling, D.~C. 2012, The Astrophysical Journal, 757, 104

\bibitem[{Robinson \& Catling(2014)}]{robinson2014b}
---. 2014, Nature Geoscience, 7, 12

\bibitem[{Seager \& Deming(2009)}]{seager2009}
Seager, S., \& Deming, D. 2009, The Astrophysical Journal, 703, 1884

\bibitem[{Selsis {et~al.}(2011)Selsis, Wordsworth, \& Forget}]{selsis2011}
Selsis, F., Wordsworth, R.~D., \& Forget, F. 2011, Astronomy and Astrophysics,
  532, 1

\bibitem[{Showman {et~al.}(2010)Showman, Cho, \& Menou}]{showman2010}
Showman, A.~P., Cho, J. Y.-K., \& Menou, K. 2010, in Exoplanets, ed. S.~Seager
  ({University of Arizona Press}), 471--516

\bibitem[{Showman {et~al.}(2015)Showman, Lewis, \& Fortney}]{showman2015}
Showman, A.~P., Lewis, N.~K., \& Fortney, J.~J. 2015, The Astrophysical
  Journal, 801, 95

\bibitem[{Showman \& Polvani(2011)}]{showman2011}
Showman, A.~P., \& Polvani, L.~M. 2011, Astrophysical Journal, 738,
  doi:10.1088/0004-637X/738/1/71

\bibitem[{Showman {et~al.}(2013)Showman, Wordsworth, Merlis, \&
  Kaspi}]{showman2013b}
Showman, A.~P., Wordsworth, R.~D., Merlis, T.~M., \& Kaspi, Y. 2013, in
  Comparative {{Climatology}} of {{Terrestrial Planets}}, ed. S.~J. Mackwell,
  A.~A. Simon-Miller, J.~W. Harder, \& M.~A. Bullock, Space Science Series
  ({University of Arizona Press}), 277--326

\bibitem[{Sobel {et~al.}(2001)Sobel, Nilsson, \& Polvani}]{sobel2001}
Sobel, A.~H., Nilsson, J., \& Polvani, L.~M. 2001, Journal of the Atmospheric
  Sciences, 58, 3650

\bibitem[{Voigt {et~al.}(2012)Voigt, Held, \& Marotzke}]{voigt2012}
Voigt, A., Held, I.~M., \& Marotzke, J. 2012, Journal of the Atmospheric
  Sciences, 69, 116

\bibitem[{Wordsworth(2015)}]{wordsworth2015}
Wordsworth, R. 2015, The Astrophysical Journal, 806, 180

\bibitem[{Yang \& Abbot(2014)}]{yang2014}
Yang, J., \& Abbot, D.~S. 2014, The Astrophysical Journal, 784, 155

\bibitem[{Yang {et~al.}(2013)Yang, Cowan, \& Abbot}]{yang2013}
Yang, J., Cowan, N.~B., \& Abbot, D.~S. 2013, The Astrophysical Journal
  Letters, 771, L45

\bibitem[{Zalucha {et~al.}(2013)Zalucha, Michaels, \&
  Madhusudhan}]{zalucha2013}
Zalucha, A.~M., Michaels, T.~I., \& Madhusudhan, N. 2013, Icarus, 226, 1743

\end{thebibliography}

\end{document}